\documentclass{ragtime}
\pdfoutput=1
\usepackage{graphicx}
\usepackage{amssymb}

\usepackage{array}

\providecommand{\lamp}{{\rm l}}
\providecommand{\newton}{{\rm n}}

\providecommand{\inc}{{\rm i}}
\providecommand{\rbreak}{r_{\rm b}}
\providecommand{\shift}{{\rm s}}
\providecommand{\bend}{{\rm b}}
\providecommand{\aber}{{\rm a}}
\providecommand{\out}{{\rm out}}
\providecommand{\dif}{\mathrm{d}} 
\providecommand{\oder}[2]{\frac{\dif #1}{\dif #2}} 

\title[The relativistic iron line in the lamp-post geometry]%
{An XSPEC model to explore spectral features from black-hole sources --- 
II.\\{\Large The relativistic iron line in the lamp-post geometry}}

\author[M. Dov\v{c}iak et al.]
       {Michal Dov\v{c}iak\at{1,a}
        Ji\v{r}\'{\i} Svoboda\at{1}
        Ren\'{e} W. Goosmann\at{2}\splitauthors
        Vladim\'{\i}r Karas\at{1}
        Giorgio Matt\at[]{3}
        and Vja\v{c}eslav Sochora\at[]{1}\\
        \ins{1}Astronomical Institute, Academy of Sciences of the Czech 
               Republic,\splitins[1]
               Bo\v{c}n\'{\i}~II 1401, CZ-141\,00~Prague, Czech Republic\\
        \ins{2}Observatoire Astronomique de Strasbourg,\splitins[2] 
               11 rue de l'Universit\'e, F-67000 Strasbourg, France\\
        \ins{3}Dipartimento di Matematica e Fisica, 
               Universit\`a degli Studi ``Roma Tre'',\splitins[3]
               Via della Vasca Navale 84, I-00146~Roma, Italy\\
        \ins{a}\Email{dovciak@asu.cas.cz}}

\begin{document}

\begin{abstract}
In X-ray spectra of several active galactic nuclei and Galactic black hole 
binaries a broad relativistically smeared iron line is observed. This feature
arises by fluorescence when the accretion disc is illuminated by hot corona 
above it. Due to central location of the corona the illumination and thus also 
the line emission decrease with radius. It was reported in the literature that 
this decrease is very steep in some of the sources, suggesting a highly compact
corona.

We revisit the lamp-post setup in which the corona is positioned
on the axis above the rotating black hole and investigate to what extent the
steep emissivity can be explained by this scenario. We show the contributions of 
the relativistic effects to the disc illumination by the primary source --- 
energy shift, light bending and aberration. The lamp-post radial illumination 
pattern is compared to the widely used radial broken power-law emissivity 
profile. We find that very steep emissivities require the primary illuminating 
source to be positioned very near the black hole horizon and/or the spectral 
power-law index of the primary emission to be very high. The broken power-law 
approximation of the illumination can be safely used when the primary source
is located at larger heights. However, for low heights the lamp-post 
illumination considerably differs from this approximation.

We also show the variations of the iron line local flux over the disc  
due to the flux dependence on incident and emission angles. The former depends 
mainly on the height of the primary source while the latter depends on the 
inclination angle of the observer. Thus the strength of the line varies 
substantially across the disc. This effect may contribute to the observed 
steeper emissivity.\footnote{This paper summarises the work done for 
the workshop {\em Ragtime12} held in 2010.}
\end{abstract}


\section{Introduction}\label{intro}

The broad iron line in the X-ray spectra of active galactic nuclei (AGN) and
Galactic X-ray binaries has been studied by various authors for more than two 
decades. The first mention of the relativistic broadening of spectral lines due 
to high orbital velocities of the accretion disc, where the iron K$\alpha$ line 
arise by fluorescence, dates as far as 1989 when \citeauthor{Fabian1989} studied
the X-ray spectrum of Cygnus X-1 observed by EXOSAT in 1983. Since then the 
relativistically broadened line was discovered in systems with diverse masses: 
in AGN with central supermassive black holes \citep[e.g.][]{Risaliti2013}, 
in X-ray binaries with the black hole of several solar masses 
\citep[e.g.][]{Miller2013} and even in systems with a neutron star 
\citep[e.g.][]{Cackett2013}. 
Although the broad lines seem very well established, one should mention that 
there exist an alternative explanation of the phenomena --- partially covering 
scenario proposed by \cite{Miller2013a}. However, recent X-ray reverberation 
studies of AGN support the reflection scenario, see e.g. \cite{Fabian2013}.

The shape of the observed line is determined by several factors: by the geometry
of the illuminating and reflecting region, by the physical properties of the
re-processing matter and by the properties of the central gravitating body.
The shaping of line, mainly its broadening, by the black-hole gravitation acting
on photons emitted in the inner accretion disc is used to measure the spin of
the black hole. Actually, high energy redshift due to large gravity near the
centre is completely responsible for the extreme width of the line. The
other components can modify the overall line profile, yet, they are not able
to change the width of the line by themselves. Still their contribution may be
important in determining the spin value. This is caused by the particular shape
of the relativistically broadened line --- the line flux gradually decreases
with the decreasing energy, thus the lower edge of the line is not easily
pinpointed, its determination depends on how strong the line's red wing is and 
the contribution of the mentioned components may be important.

One of the inevitable components, that makes the formation of the fluorescent
line possible, is the illuminating corona. Its geometry will affect the
illumination of the disc and consequently also the emission of the line from
different parts of the disc. This will eventually alter the overall line
profile. Usually the corona is supposed to be either extended 
\citep[e.g.][]{Wilkins2012} over large area above the disc or concentrated in 
a compact region \citep[e.g.][]{Fabian2012}.
In the first case the illumination of the disc is often assumed to be a broken
power-law function of the radius, with more intensive illumination and resulting
higher line emission in the inner parts of the disc. Sometimes the observed
radial power-law near the black hole is quite steep \citep{Wilkins2011} and it 
was suggested \citep{Svoboda2012a} that it could be caused by the second 
possible scenario, i.e. sort of a lamp-post configuration, where the compact 
patch of the corona located above the black hole illuminates the accretion disc,
sometimes referred to as an aborted jet scenario \citep{Ghisellini2004} or 
a light bending model \citep{Miniutti2004}. 
In this geometry, the illumination of the disc is due to a compact
primary source and photon trajectories close to the centre are bent by
strong gravity of the black hole. Consequently, the radial profile of the line
emission takes a particular form that depends on the height of the lamp-post.

In this paper we compare the two radial profiles of the line
emission --- the broken power-law dependence and illumination in the lamp-post
scenario. We concentrate mainly on the question if the observed
steep radial decrease of the emissivity could be interpreted in terms of the
lamp-post geometry. To this purpose we at first assume very simple local 
physics of the emission, particularly, the emission does not depend on incident 
and emission angles, and the flux in line is isotropic and proportional
to the incident flux. Then, we also apply the emission directionality given by
the numerical modelling of radiative transfer \citep[using Monte Carlo 
multi-scattering code NOAR, see][]{Dumont2000}.


\section{Relativistic lamp-post geometry}\label{lamp}

The lamp-post geometry has been introduced to describe the observed emission
from X-ray irradiated accretion discs by \cite{Matt1991} and 
\cite{Martocchia1996}. The model
consists of an X-ray source (`lamp') producing the primary irradiation and
representing an optically thin corona that is thought to extend above the
optically thick medium of a standard accretion disc \citep{Frank2002}. This 
scheme has proved to be very popular in the context of accreting supermassive 
black holes in cores of AGN \citep{Peterson1997}. Location of the primary source 
on the black hole axis can be imagined, e.g., as a site of action where jets are
initially accelerated \citep{Biretta2002} or where the shocks in an aborted jet
collide \citep{Ghisellini2004}. A down-scaled version of the model has been also
invoked to describe microquasars \citep{Mirabel1998}.

The lamp-post geometry (on or off-axis) has already been studied in various 
context by several authors -- AGN variability was studied by 
\citep{Miniutti2004} and \cite{Niedzwiecki2010}, the polarisation properties 
were investigated by \cite{Dovciak2011a} and the X-ray reverberation mapping by
\cite{Emmanoulopoulos2014} and \cite{Cackett2014}. Recently, \cite{Dauser2013} 
has studied disc reflection due to illumination by a jet, i.e. radially extended 
region moving along the axis.

Despite the fact that realistic corona must be a very complex, inhomogeneous
and turbulent medium, the lamp-post model captures the main components of a
typical AGN spectrum, and it allows us to search for the parameter values. In
particular, the slope of the primary power-law continuum, and the skewed and 
redshifted profile of the broad iron line around 6--7 keV that has been 
interpreted in terms of relativistically smeared reflection spectrum.

It has been shown \citep{Wilms2001} that a steep emissivity profile of
$\simeq4.3$-–$5.0$ of the iron-line and reflection features are required in
XMM-Newton observation of MCG--6-30-15. This has been interpreted in terms of
highly central concentration of the irradiating flux, in a much more compact
nuclear region than predicted by pure accretion disc models. Similarly
steep emissivity profile has been reported in 1H0707-495 \citep{Fabian2009} 
and IRAS13224-3809 \citep{Ponti2010}. In
order to explain the unusually steep spectrum, \citet{Wilms2001} invoke some
additional X-ray source that is presumably associated with the extraction of the
black hole spin energy, perhaps via some kind of magnetic coupling
\citep{Blandford1977}.

The main aim of the present investigation is to verify whether the relativistic
effects can produce the steep emissivity required by the mentioned observations.
To this end we consider Kerr metric for the gravitation of a rotating black
hole, and we allow for both prograde and retrograde rotation of the accretion
disc with respect to the black hole spin.

A complex interplay of the energy shifts, aberration, boosting and light-bending
effects acts on the primary as well as reflection components of the X-ray
spectrum, especially when the source of irradiation is placed at a small height
near above the horizon and if the black hole rotates rapidly, so that the inner
edge of the disc is at a small radius. As a result of this interplay, it is
not obvious at all whether the resulting emissivity comes out significantly
steeper in comparison to the non-relativistic limit of an irradiated standard
disc.

To study the radial emissivity we first turn our attention to the reflected
line component. We assume the line flux to be proportional to the incident
flux and the photons will be emitted isotropically in local frame co-rotating
with the Keplerian disc, no matter what the incident and emission angles are.
In this way we are going to study the effect of the relativistic lamp-post
geometry only, separating it from the effects due to the dependence of the local
physics on geometry of the incident and emission light rays. 

In this approach the local line emission in the disc is proportional to
the normalization of the power-law incident flux. We assume the primary emission
to be isotropic in local frame and that it is a power law with the photon
index $\Gamma$, i.e. $f_\lamp(E_\lamp)=N_\lamp E_\lamp^{-\Gamma}$. Then the
incident flux, $f_\inc$, is a power law with the same photon index but with a
different normalization
\begin{equation}
\label{eq:incident_flux}
f_\inc(E_\inc)=N_\inc(r) N_\lamp E_\inc^{-\Gamma}\ .
\end{equation}
Here, the normalization $N_\inc(r)$ is given by the curved geometry of the
light rays and relativistic shift of the energy. It can be expressed in the
following way \citep[see e.g.][]{Dovciak2004d}
\begin{equation}
\label{eq:N_inc1}
N_\inc(r)=g_\inc^\Gamma\,\oder{\Omega_\lamp}{S_\inc}=
\frac{g_\inc^{\Gamma-1}}{U^t_\lamp}\,\oder{\Omega_\lamp}{S}\ .
\end{equation}
The primary photons emitted by the lamp-post into the local solid angle
$\dif\Omega_\lamp$ fall down onto the disc area measured in the frame co-moving
with the disc $\dif S_\inc=p_{\inc\mu} U^\mu \,\dif S=g_\inc\,U_\lamp^t\,\dif S$.
These photons are shifted to the incident energy $E_\inc$ from the emission
energy $E_\lamp$ by the energy shift
$g_\inc=E_\inc/E_\lamp=p_{\inc\mu}U^\mu/p_{\lamp\mu}U_\lamp^\mu$ which is
responsible for the factor of $g_\inc^\Gamma$ in the above equation. We have
denoted the four-momentum of the incident photons by $p_\inc^\mu$, the
four-velocity
of the static lamp-post by $U^\mu_\lamp=(U^t_\lamp,0,0,0)$ and the four-velocity
of the disc by $U^\mu$. We assume the disc to be Keplerian above the marginally
stable orbit and freely falling below it with the constant energy and momentum
that the matter had at this orbit. The area element $\dif S=r\dif r\dif\varphi$
is evaluated in Boyer-Lindquist coordinates.

The normalization of the incident flux, $N_\inc$, is a function of radius, and
thus it determines the radial emission profile of the line flux. We can
separate this function into several components
\begin{equation}
\label{eq:N_inc2}
N_\inc(r)=\frac{1}{r}\,
\oder{\mu_\newton}{r}\,\times\,\frac{g_\inc^{\Gamma-1}}{U^t_\lamp}\,
\,\times\,\oder{\mu}{\mu_\newton}\,\times\,
\oder{\mu_\lamp}{\mu}\ .
\end{equation}

\begin{figure}
\centering
\includegraphics[width=\textwidth]{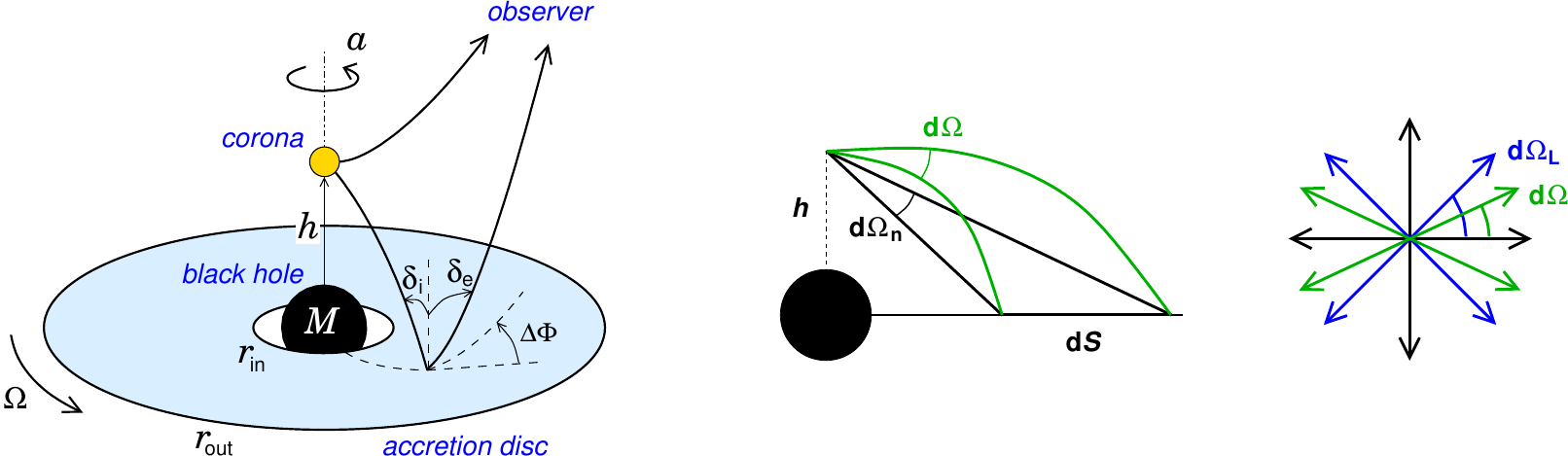}
\caption{{\em Left:} The sketch of the lamp-post geometry.
{\em Middle:} Due to the light bending the photons illuminating the same
area of the disc are emitted by the primary source into different solid angles
in the relativistic and the Newtonian cases.
{\em Right:} The photons emitted isotropically in the rest frame of the primary
source are beamed perpendicularly to the rotation axis because of the strong
gravity near the black hole.}
\label{fig:dWadSd}
\end{figure}

The first component is chosen in such a way that it represents exactly the
Newtonian value of $N_\inc(r)$
\begin{equation}
\label{eq:N_inc_newton}
N_\inc^{\newton}(r)\equiv\oder{\Omega_\newton}{S}=\frac{1}{r}\,
\oder{\mu_\newton}{r}=\frac{h}{(r^2+h^2)^{3/2}}\ .
\end{equation}
In the above, we have introduced the Newtonian angle of emission,
$\theta_\newton$, as the angle under which the primary photon has to be emitted
from the lamp-post at height $h$ in the Newtonian non-curved space so that it
falls down onto the disc at the
radius $r$ and $\mu_\newton\equiv\cos{\theta_\newton}$.

The second component, $g_\inc(r)^{\Gamma-1}/U_\lamp^t$, is connected with the
energy shift of the incident photons and it should be emphasized that it depends
on the primary flux via the photon index $\Gamma$. For Keplerian discs in the
Kerr space-time it can be expressed above the marginally stable orbit as
\begin{equation}
\label{eq:N_inc_shift}
N_\inc^{\shift}(r)=
\left(\frac{r^2+a\sqrt{r}}{r\sqrt{r^2-3r+2a\sqrt{r}}}\right)^{\Gamma-1}\,
\left(1-\frac{2h}{h^2+a^2}\right)^{\frac{\Gamma}{2}}\ .
\end{equation}

The third component,
\begin{equation}
\label{eq:N_inc_bend}
N_\inc^{\bend}(r)\equiv\oder{\Omega}{\Omega_\newton}=
\oder{\mu}{\mu_\newton}=
\frac{\sin{\theta}}{\sin{\theta_\newton}}\,
\oder{\theta}{\theta_\newton}\ ,
\end{equation}
represents the effects of the light bending in the
curved space-time. It compares the solid angle $\dif\Omega$ in the
Boyer-Lindquist coordinates with the Newtonian value, defined above, into which
primary photons have to be emitted to illuminate the disc area $\dif S$ at the
disc radius $r$.

The fourth component represents the ``gravitational aberration''. Due to the
fact that the local observers on the axis measure the distances differently
along the axis and perpendicular to it, the local isotropic emission will be
beamed in Boyer-Lindquist coordinates in the direction perpendicular to the
axis. We can express it by comparing the solid angle in local frame of the
lamp-post with the solid angle in Boyer-Lindquist coordinates
\begin{equation}
\label{eq:N_inc_aber}
N_\inc^{\aber}(r)=
\oder{\Omega_\lamp}{\Omega}=\frac{h}{\sqrt{\Delta_{\rm h}}}
\left[1+\left(\frac{\Delta_{\rm h}}{h^2}-1\right)
\cos^2{\theta_\lamp}\right]^{3/2}\ ,
\end{equation}
where $\Delta_{\rm h}\equiv h^2-2h+a^2$ and the photon's local emission angle
$\theta_\lamp$ is a function of the radius $r$ at which such a photon strikes
the disc. One can see that the solid angle $\dif\Omega$ is amplified by the
factor $h^2/\Delta_{\rm h}>1$ along the axis ($\theta_\lamp=0^\circ$ or
$\theta_\lamp=180^\circ$) and it is diminished by the factor
$\sqrt{\Delta_{\rm h}}/h<1$ in the direction perpendicular to the axis
($\theta_\lamp=90^\circ$).

\begin{figure}
\centering
\includegraphics[width=\textwidth]{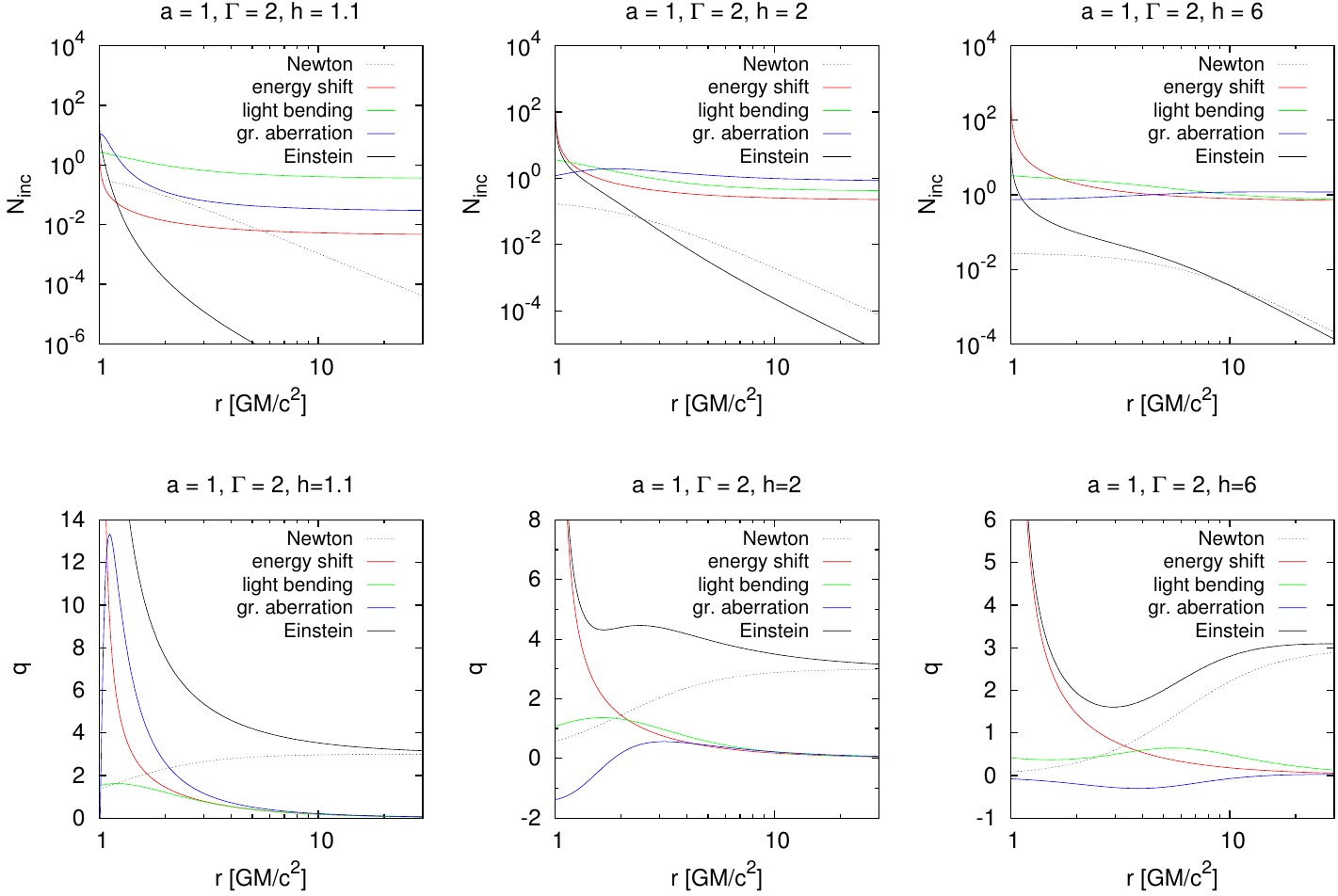}
\caption{The radial profile of the incident flux $N_\inc(r)$ (top) and its
radial power-law index $q(r)$ (bottom) in the relativistic lamp-post geometry
with the illuminating
primary source at heights $h=1.1,\,2$ and $6\,GM/c^2$ (from left to right).
The primary flux is a power law with the index $\Gamma=2$ and the Kerr black
hole rotates extremely with the spin $a=J/M=1\,GM/c$. The contributions from the 
Newtonian, energy shift, light bending and 
gravitational aberration parts are shown. The solid black line (Einstein) 
depicts the overall incident flux and its radial power-law index.}
\label{fig:Ninc_power_parts}
\end{figure}
On the top panels of Fig.~\ref{fig:Ninc_power_parts} we show the radial profile
of the function $N_\inc(r)$ and its components for extremely rotating black hole
and for several heights of the primary source. In Newtonian case the
illumination of the disc is flat below the lamp-post and decreases with the
radius with the third power far from the centre. The energy shift component is
higher than unity for the radius lower than the height of the lamp-post
(however, due to the Doppler shift and black hole spin the transition radius is
somewhat shifted)
and it is lower than unity above this radius. It is due to the fact that in the
first case the photon falls closer to the black hole, gaining the energy,
whereas in the second case it climbs out of the gravitational potential well,
losing its energy. As a result, the shift gains very high values for small
radii close to the horizon and quite low values far from the black hole if the
height of the primary source is low.

The effect of light bending is stronger closer to the black hole. Thus the
photon trajectory that is nearer to the black hole is more curved, the
difference in bending of two close trajectories gets smaller farther away from
the centre. This results in light bending component $N_\inc^{\bend}(r)$ to be a
decreasing function of radius.

Due to the fact that the gravitational aberration decreases the solid angle,
which photons are emitted into, the most in the direction perpendicular to the
axis, the incident flux will be amplified for those radii where the photons
emitted in this direction strike the disc. That is why the component
$N_\inc^{\aber}(r)$ first increases with the radius and then decreases.
The maximum moves farther away from the black hole for higher lamp-post. For
very low heights of the primary source, the photon trajectories emanating
perpendicularly to the axis are bend so much that they do not strike the disc,
rather they fall onto the horizon. In that case this component of the incident
flux decreases with the radius.

To compare the relativistic lamp-post illumination with the broken power law
one, we define the radial power-law index for the lamp-post geometry as the
slope of the radial profile of $N_\inc(r)$ in the log-log graph (i.e. slope of
the graphs on top panels in Fig.~\ref{fig:Ninc_power_parts}). The definition
reads
\begin{equation}
\label{eq:q}
q(r)\equiv \oder{\log{N_\inc(r)}}{\log{r}} =
-r\oder{}{r}\ln{N_\inc}(r)\ .
\end{equation}
The radial power-law index defined in this way clearly depends on the radius.
The four components of the incident flux that has to be multiplied to give the
overall illumination translate into four components of the power-law index
$q(r)$ that have to be added to give the overall relativistic radial power-law
index. We show all four components of the index on bottom panels in
Fig.~\ref{fig:Ninc_power_parts}.

The index $q(r)$ for large radius is given by the Newtonian value,
$q(r\rightarrow\infty)=3$. For very low radii, close to the horizon, the energy
shift component dictates the behaviour of the $q(r)$. Its influence extend
farther for higher primary spectral power-law index. For very low heights,
heights near above the black hole horizon, the component due to gravitational
aberration adds also quite significantly to the index for low radius. For
low lamp-post heights ($h\lesssim 6\,GM/c^2$), the light bending and
gravitational aberration component may create local maxima in the radial
power-law index, depending on the black hole spin and primary spectral power-law
index $\Gamma$. More examples of the behaviour of the radial power-law index
$q(r)$ for different parameter values are shown in Fig.~\ref{fig:power}. Note,
that if the line emission were proportional to the incident flux also below the
marginally stable orbit (dotted lines in Fig.~\ref{fig:power}), the index $q(r)$
for lower absolute value of the black hole spin, i.e. larger radius of the event
horizon, would be larger. This is due to the energy shift component of the
$q(r)$, which gains large values also at higher radii.

From Figs.~\ref{fig:Ninc_power_parts} and \ref{fig:power} it is evident that the
relativistic lamp-post illumination is very different from the broken power law,
which would be represented by two constant values in these figures.
For comparison, we show the relativistic lamp-post emission together with
a broken power law in Fig.~\ref{fig:Ninc}. The graphs in this figure are
renormalized in such a way that they do not intersect each other, here, we are
interested in their shape only. The broken-power-law graphs (depicted by red)
have Newtonian value of the index, $q_\out=3$, above the break radius,
$\rbreak$. The break radius and index $q$
below it were chosen by eye so that they approximately represent the 
relativistic lamp-post flux. We show their values in
Tables~\ref{tab:q1}--\ref{tab:q4}. One can see that the broken power law is 
close enough only in some radial regions whereas it fails for small radii near
the horizon and region around the break radius. Although the difference between
the two is large in these regions (note, that the graphs are in logarithmic
scale), one still cannot jump to the conclusion that the broken power-law 
approximation would fail in fitting the spectra
originated in the relativistic lamp-post geometry. The energy of the photons
coming to the observer from regions close to the horizon is strongly shifted
to very low values and the troublesome region near the break radius may be small 
enough with respect to the whole disc to change the overall spectrum. Thus the
spectra for the broken power-law emissivity and for relativistic lamp-post 
geometry might still be similar sufficiently.

From graphs in Fig.~\ref{fig:Ninc} one can see that the illumination profile 
would actually be much better approximated with a power-law with two breaks 
instead of one, especially for higher locations of the primary source. 
Comparisons between simple power-law and once or twice broken power-law are
investigated by \cite{Wilkins2011}.

\section{The directionality of the local flux}

The flux emitted by the primary source illuminates the disc and the incident
photons are then re-processed in the orbiting material. They scatter on 
electrons, are absorbed by ions or neutral atoms or they can be created
by the fluorescence when electrons in ions or neutral atoms change their state.
Fluorescent spectral lines, line edges and Compton hump are typical features of
such reflected X-ray spectra \citep{Ross2005, Garcia2013}. The most prominent
spectral line in this energy band is that of iron (Fe K$\alpha$ line doublet for 
neutral iron is at $6.4\,$keV) due to its large abundance and high fluorescence 
yield. 
The flux emitted locally in this line depends on number of absorbed photons that
create the vacancies and fluorescent yield which characterises how fast these 
vacancies fill. If we assume that there is always enough photons that induce the 
fluorescence then the flux in the line is mainly dependent on the absorption. 
A vacancy at the K level of a neutral iron line is created when a photon with
the energy above the iron K edge (at approx. $7.1\,$keV) is absorbed. The 
efficiency of the absorption quickly decreases with the energy, thus only 
photons up to a few keV above this edge are absorbed. 
This is due to the fact that the K-absorption cross-section of a photon with 
energy $E$ above the K-absorption edge at 7.1 keV (measured in the local disc 
frame) decreases approximately as \citep{Verner1993}
$1.9\,(E / 7.1\,{\rm keV})^{-3.1} - 0.9\,(E / 7.1\,{\rm keV})^{-4.1}$ and thus 
levels off to 1\% of its initial value already at $37.4\,$keV. Since the primary 
spectrum assumed in our model 
extends to much higher energies, the flux in the line is simply proportional to 
the normalisation of the incident power-law spectrum. We generally assume that 
both cut-off energies of the primary spectrum lie outside the energy band where 
absorption occurs. We give two examples for extremely rotating Kerr black hole 
to show how well this assumption is fulfilled:

\begin{enumerate}
\item The lower energy cut-off is shifted to higher energy when the primary
      source is very high above the disc. Then the incident photons gain the
      highest energy if they fall close to the horizon. For the lamp at height 
      $h=100\,$GM/c$^2$ and incident radius at $r_{\rm i}=1.035\,$GM/c$^2$ the
      photon energy shift is $g_{\rm i}=67$. Thus the lower energy cut-off at
      $0.1\,$keV would be shifted to $6.7\,$keV which is still below the Fe
      K edge. Note, that the emission below this region will have very low 
      contribution to the overall spectral shape of the observed broadened line
      both due to small emission area and due to small value of the transfer 
      function (that amplifies local flux when transferred to the observer at 
      infinity), $G<0.1$ for inclination $\theta_{\rm o}=70^\circ$ (and smaller
      for lower inclinations). The contribution from this region will be shifted
      by the factor $g<0.1$, thus to energy $E<0.64\,$keV.
\item The high energy cut-off is shifted to lower energy when the primary source
      is very low above the black hole horizon. Then the incident photons lose
      the energy when they have to climb out of deep potential well, thus they
      lose more if they fall to the disc far away from the horizon.
      For the lamp at height $h=1.3\,$GM/c$^2$ and incident radius at 
      $r_{\rm i}=1000\,$GM/c$^2$ the photon energy shift is $g_{\rm i}=0.186$. 
      Thus the higher energy cut-off at $200\,$keV would be shifted to 
      $37.2\,$keV which is still high enough above the Fe K edge. Note, that the 
      emission above this region will have quite low contribution to the overall 
      spectral shape of the observed broadened line due to radial decrease of
      the line emissivity as $r^{-3}$.
\end{enumerate}

We have computed the reflection from a neutral disc in constant density slab
approximation by the Monte Carlo code NOAR \citep{Dumont2000}. The line flux was
then computed by subtracting the interpolated reflected continuum from
the reflection spectra. The line flux includes also the Compton shoulder created 
by scattering of the fluorescent photons before they leave the disc. 
We approximate the line with a narrow box function with a width of $1\,$eV 
(simulating a delta function) that has the numerically computed flux. 
This speeds up the code without loss of precision since the 
relativistically broadened line does not depend on the exact shape of the 
locally narrow line. The local Fe K$\alpha$ flux depends on 
incident and emission angles due to the fact that incident photon travels
different distances in different layers during radiative transfer in the disc.
On the other hand it does not depend on the azimuthal angle between incident 
and emitted light rays. We define the emission directionality function as the 
numerically computed flux in line per unit normalisation of the incident 
power-law flux 
\begin{equation}
\label{eq:M}
{\cal M}(\mu_{\rm i},\mu_{\rm e})\equiv\frac{\dif{N}}{\dif{t}\,\dif{S^\perp}
\dif{\Omega}\,\dif{E}}=\frac{1}{2\pi\,\mu_{\rm e}}\frac{\Delta N}
{\Delta\mu_{\rm e}\,\Delta{E}\,N_{\rm tot}}\int_{E_0}^{E_{\rm c}}{E^{-\Gamma}}
\dif{E}\ ,
\end{equation}
where $\Delta N$ is the number of photons emitted into the emission angle bin 
characterised by its cosine, $\Delta\mu_{\rm e}$, i.e. into the whole azimuth 
of $2\pi$, hence the leading factor in the definition, and into energy bin 
$\Delta E$. $N_{\rm tot}$ is the total number of photons used in Monte Carlo 
computation and thus we multiply by the integrated energy dependence to 
normalise it as mentioned earlier, i.e. for incident power-law being 
exactly $F_\inc(E)=E^{-\Gamma}$. In the definition (\ref{eq:M}) there is one 
more factor of $1/\mu_{\rm e}$
due to the local flux being defined with respect to the area perpendicular to 
the emitted light ray while the reflected number of photons was computed per 
unit disc area. The sharp low, $E_0$, and high, $E_{\rm c}$, energy cut-offs at 
$2$ and $300\,$keV, respectively, were used in the computation. 
With this definition of emission directionality, the local line flux 
is defined as
\begin{equation}
\label{eq:local_flux}
F_{\rm loc}(E) \equiv {\cal R}(r)\, {\cal M}(\mu_{\rm i}, \mu_{\rm e})\,
\delta(E-E_{\rm rest})
\end{equation}
with $E_{\rm rest}=6.4\,$keV being the rest energy of the neutral Fe K$\alpha$ 
line and radial dependence of the normalisation of the incident power-law as 
discussed in the previous section, ${\cal R}(r)=N_\lamp\,N_{\rm i}(r)$, see 
eq.~(\ref{eq:incident_flux}).

\begin{figure}[t]
\centering
\includegraphics[width=0.95\textwidth]{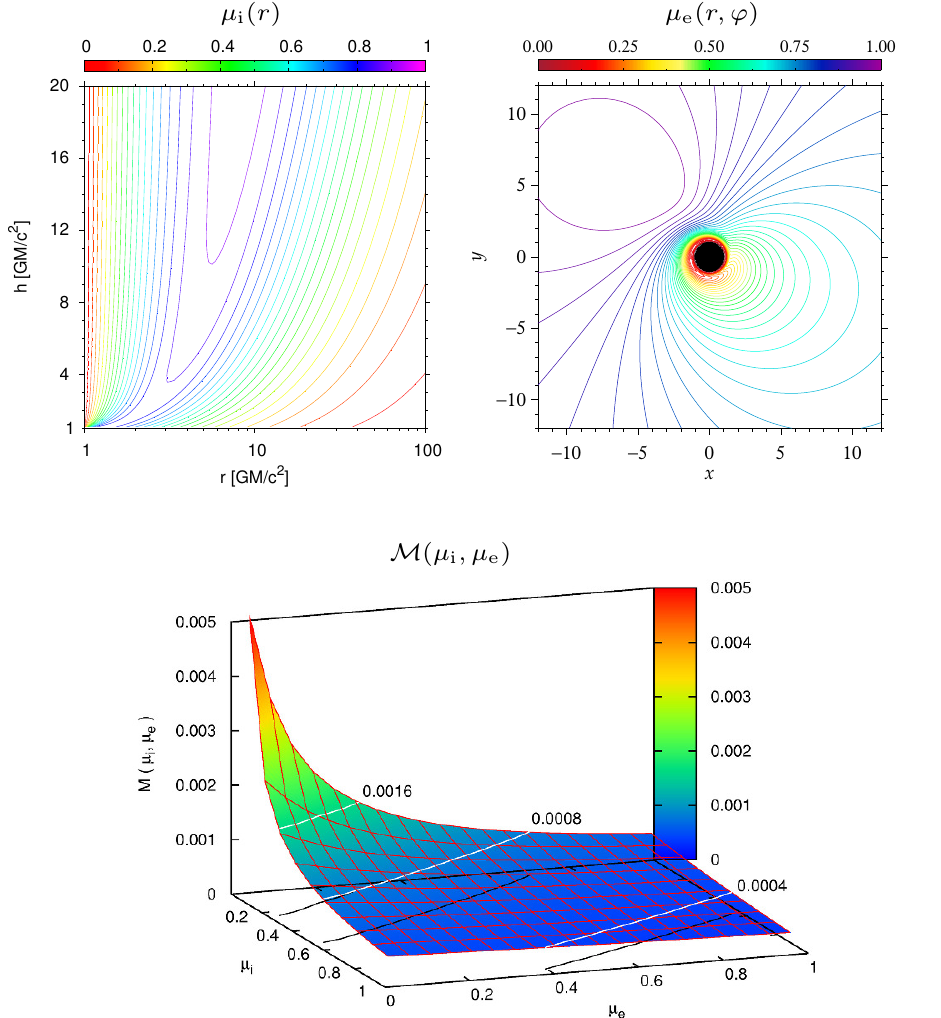}
\caption{\label{fig:local_flux} {\em Top left:} The radial dependence of the 
cosine of the incident angle, $\mu_{\rm i}$, for different heights of the 
primary source and for the extremely rotating black hole with the spin 
$a=1\,GM/c$. {\em Top right:} The dependence of the cosine of the emission 
angle, $\mu_{\rm e}$, on the position on the disc. The spin of the black hole is 
$a=1\,GM/c$ and the inclination of the observer is $\theta_{\rm o}=30^\circ$.
{\em Bottom:} The emission directionality function, 
$M(\mu_{\rm i},\mu_{\rm e})$, is depicted for the photon index $\Gamma=2$. 
Notice the high values it acquires for large incident and emission angles.}
\end{figure}

We show the emission directionality function, $M(\mu_{\rm i},\mu_{\rm e})$, in 
the bottom panel of Fig.~\ref{fig:local_flux} for the photon index of the 
primary
radiation $\Gamma=2$. To see which values this function may acquire we also
show the values of cosines of incident and emission angles at the top panel of 
the same figure (maps of cosine of emission angles are also shown in 
Figs.~\ref{fig:gcosG_100}, \ref{fig:gcosG_200} and \ref{fig:gcosG_300}). One can 
see that more radiation is emitted when the incident 
angle is large (measured from the normal to the disc), i.e. when the photons 
arrive almost parallelly with the disc. The same applies for the angular
dependence of emissivity which obeys limb brightening law. The brightening is,
however, smaller than the limb brightening law derived by \cite{Haardt1993}, 
where ${\cal M}(\mu_{\rm e})\sim {\rm ln}(1+\mu_{\rm e}^{-1})$. Since both the 
incident as well as emission angles are very high close above the horizon due to 
aberration caused by high Keplerian velocity, the emission directionality will
be highest in this region. 
We show the map of
${\cal M}(\mu_{\rm i}(r,\varphi),\mu_{\rm e}(r,\varphi))$ in the equatorial 
plane for several values of black hole spin, observer inclination and height of 
the primary source in Figs.~\ref{fig:M_100}, \ref{fig:M_200} and 
\ref{fig:M_300}. One can already appreciate the importance of the limb 
brightening effect by 
comparing the values of this function with the values of energy shift, $g$, and 
transfer function (i.e. the amplification of the local emission due to 
relativistic effects), $G$, shown in Figs.~\ref{fig:gcosG_100}, 
\ref{fig:gcosG_200} and \ref{fig:gcosG_300}. Thus we can expect that the shape 
of the broad iron line may be substantially influenced by the emission 
directionality. Moreover, the dependence of the emission
directionality on the radius through the radially dependent incident and 
emission angles might cause that the observed radial emissivity profile, 
characterised by the radial power-law index $q$, might be measured with a 
systematic error if wrong assumption on emission directionality is taken
\citep{Svoboda2014}.

\section{The shape of the relativistic line in lamp-post geometry}

In the previous two sections we have discussed the local line flux and its 
dependence on the disc illumination, that gave us the radial part of the local 
emission, and local re-processing in the disc, that determined the emission 
directionality. The final shape of the observed spectral line is influenced by 
the relativistic effects that change the spectral properties of the local 
emission when transferred to the observer at infinity. The local spectrum 
will be shifted in energy due to Doppler shift and gravitational redshift, and 
it will be amplified due to Doppler boosting, gravitational lensing, 
aberration and light bending (the last two influence the local emission angle
i.e. change the projections of the emitting area). To get the observed shape of 
the line one has to integrate the local emission over the whole disc
\begin{equation}
\label{eq:observed_line}
F_{\rm obs}(E) \equiv \frac{\dif{N_{\rm obs}}}{\dif{t}\,\dif{\Omega}\,\dif{E}} = 
\int \dif{S}\, G\,F_{\rm loc}\,\delta(E-gE_{\rm rest})\ ,
\end{equation}
where $G$ is the transfer function \citep[see e.g.][]{Cunningham1975, 
Dovciak2004d} characterising an amplification of the local line
flux, $F_{\rm loc}={\cal R}(r){\cal M}(\mu_{\rm i},\mu_{\rm e})$, which is 
shifted to the observed energy by the g-factor,
$g=E/E_{\rm rest}$. Note, that the $\delta$-function in this equation is in the 
observed energy while in the eq.~(\ref{eq:local_flux}) it was in the local
energy. The transfer function for a photon number density flux is 
$G=g^2\,l\,\mu_{\rm e}$, where the lensing, $l$, characterises amplification due 
to focusing of the light rays (caused by light bending). As mentioned in the 
previous sections, each part that contributes to the overall shape of the 
observed line, ${\cal R}(r)$, ${\cal M}(\mu_{\rm i}, \mu_{\rm e})$, 
$G(r,\varphi)$ as well as the energy shift $g(r,\varphi)$ are depicted in the
Appendices~\ref{app:N_inc} and \ref{app:maps}. Additionally we also show the 
overall map of the observed flux $F_{\rm obs}(r,\varphi)=G\,{\cal R}(r)
{\cal M}(\mu_{\rm i},\mu_{\rm e})$ in the equatorial plane in 
Figs.~\ref{fig:F_100}, \ref{fig:F_200} and \ref{fig:F_300}. Note, that in 
the eq.~(\ref{eq:observed_line}) for each observed energy one integrates this 
function along the energy shift contour.

\begin{figure}[t]
\centering
\includegraphics[width=\textwidth]{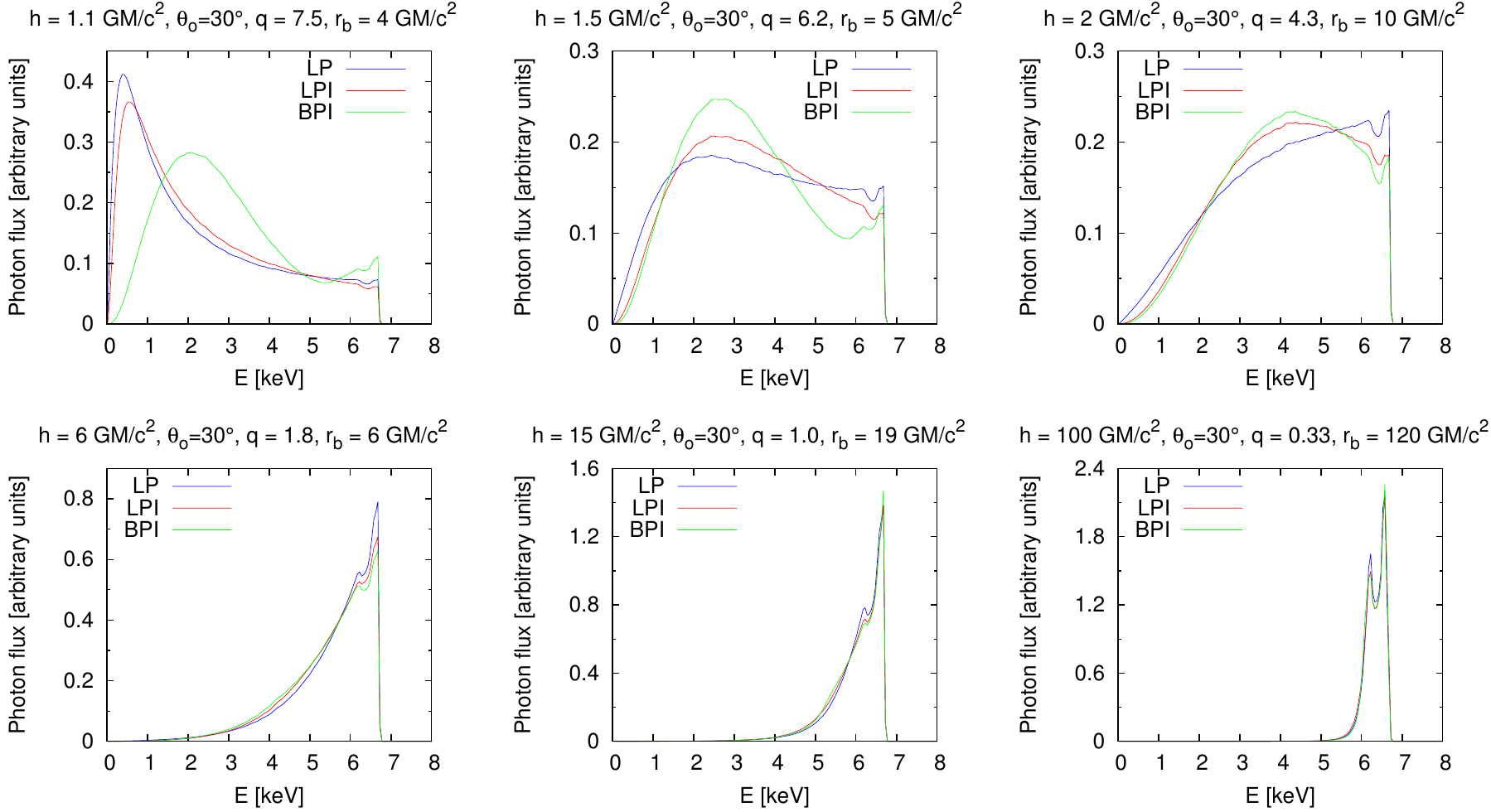}
\caption{The comparison between the shape of the line in the lamp-post geometry
with numerically computed angular directionality (blue) and with isotropic
emission (red), and with the radial broken power-law emission with isotropic
directionality (green). The height, $h$, the value of the
inner radial power-law index, $q$, and the break radius, $r_{\rm b}$, where it
changes to $q_{\rm out}=3$, are shown at the top of each graph. The inclination
of the observer is $\theta_{\rm o}=30^\circ$, the spin of the black hole is
$a=1\,GM/c$ and the photon index of the primary source is $\Gamma=2$.}
\label{fig:LPxBP_30}
\end{figure}

The shape of the relativistically broadened spectral line of iron for different
assumptions on radial emissivity and emission directionality is shown in 
Fig.~\ref{fig:LPxBP_30}. One can see that the broken power-law emissivities 
result in quite a different line shape only for a very low locations of the
corona when compared with a lamp-post illumination profile, in both cases an
isotropic local emission being assumed. The differences 
might be very well explained by comparing the emissivity profiles in 
Fig.~\ref{fig:Ninc}. The broken power-law emissivity underestimates the 
flux, the largest deficiency occurs in the region very close to the black hole, 
where the gravitational redshift is large, and around the break radius, 
$\rbreak$. Note, that in Fig.~\ref{fig:LPxBP_30}
the line flux is in all cases normalised to unit total flux, so the spectral 
line for broken power-law radial profile is not below that one for the lamp-post 
geometry for all energies. One can see, however, that the line flux is much 
lower in two energy bands, one, where the energy shift is large with small 
values of g-factor, $g\ll1$, (i.e. for low energies) and one when the g-factor is 
widely spread around unity (i.e. energies 
around iron line rest energy) that corresponds to the break radius region.
Note, that the deficiency in the flux for low heights changes to an excess in 
flux for high heights of the primary source.

Further differences in the line shape arise when isotropic emission is compared
with the numerically computed one given by the emission directionality function
${\cal M}(\mu_{\rm i}, \mu_{\rm e})$. Again these differences are large only for 
low heights of the corona. The numerically computed directionality results in
larger flux for low energies and energies around the rest energy of the line. 
This is mainly due the incident angle being very high both in the vicinity of 
the black hole as well as farther away from the centre (see the top left panel in
Fig.~\ref{fig:local_flux}) when the emission directionality function acquires
higher values (see the bottom panel in the same figure).

\begin{figure}[t]
\centering
\includegraphics[width=\textwidth]{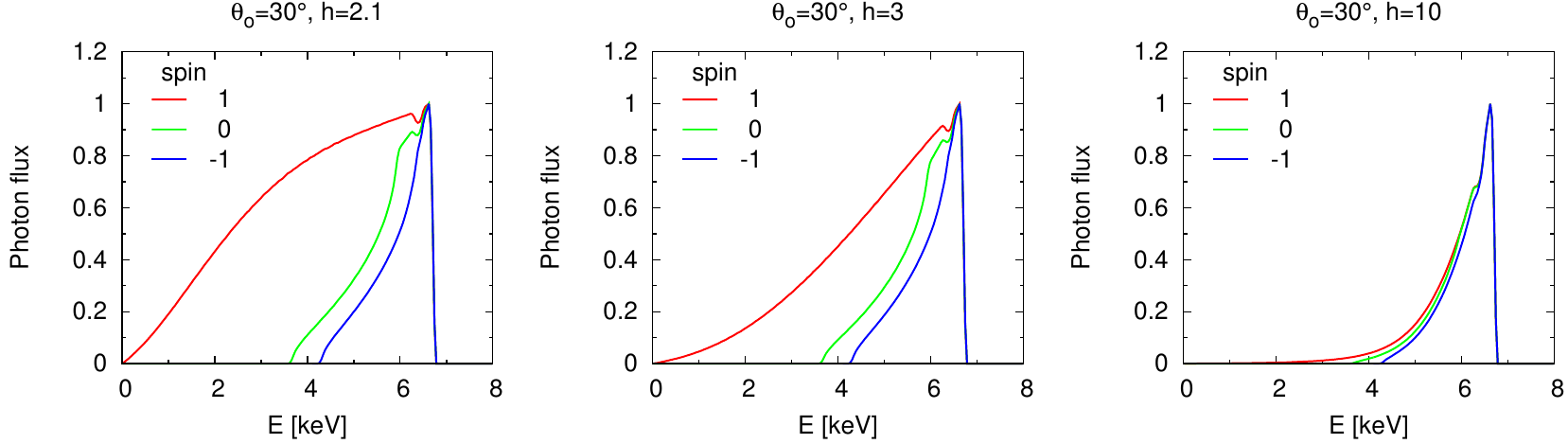}
\caption{The comparison between the shape of the line in the lamp-post geometry
for the Schwarzschild (green) and co-rotating (red) and counter-rotating (blue) 
extreme Kerr black holes for the primary source height, $h=2.1,\,3$ and 
$10\,GM/c^2$. The inclination of the observer is $\theta_{\rm o}=30^\circ$ and 
the photon index of the primary source is $\Gamma=2$.}
\label{fig:LP_30}
\end{figure}

To see how the shape of the relativistically broadened line depends on the 
height  of the source, let's compare the line for three different spins and 
three 
different heights (Fig.~\ref{fig:LP_30}). One immediately sees that the line
is much narrower for higher heights even for the extreme Kerr black hole. This 
is due to the fact that the disc is illuminated much more homogeneously from
higher lamps and since the area of the inner part of the disc, where the red 
wing
of the line arises, is very small compared to the area of the whole disc. Thus 
the shape of the line changes very little for different black hole spins if the
corona is positioned more than $10\,$GM/c$^2$ above the centre. Opposite is also 
true, i.e. if the black hole counter-rotates with an extreme spin, one would not
be able to distinguish between different heights of the corona if the height is
below approximately $10\,$GM/c$^2$ above the centre. This is due to the fact
that in this case the hole in the disc below marginally stable orbit 
($r_{\rm ms}=9\,$GM/c$^2$) is quite large and the disc illumination for small 
heights of the primary source changes mainly below this radius while it does not
change that much above the inner edge of the disc.

\section{Application to MCG-6-30-15}

\begin{figure}[t]
\centering
\parbox[t]{0.45\textwidth}
{\hspace*{-13mm}\includegraphics[width=0.5\textwidth]{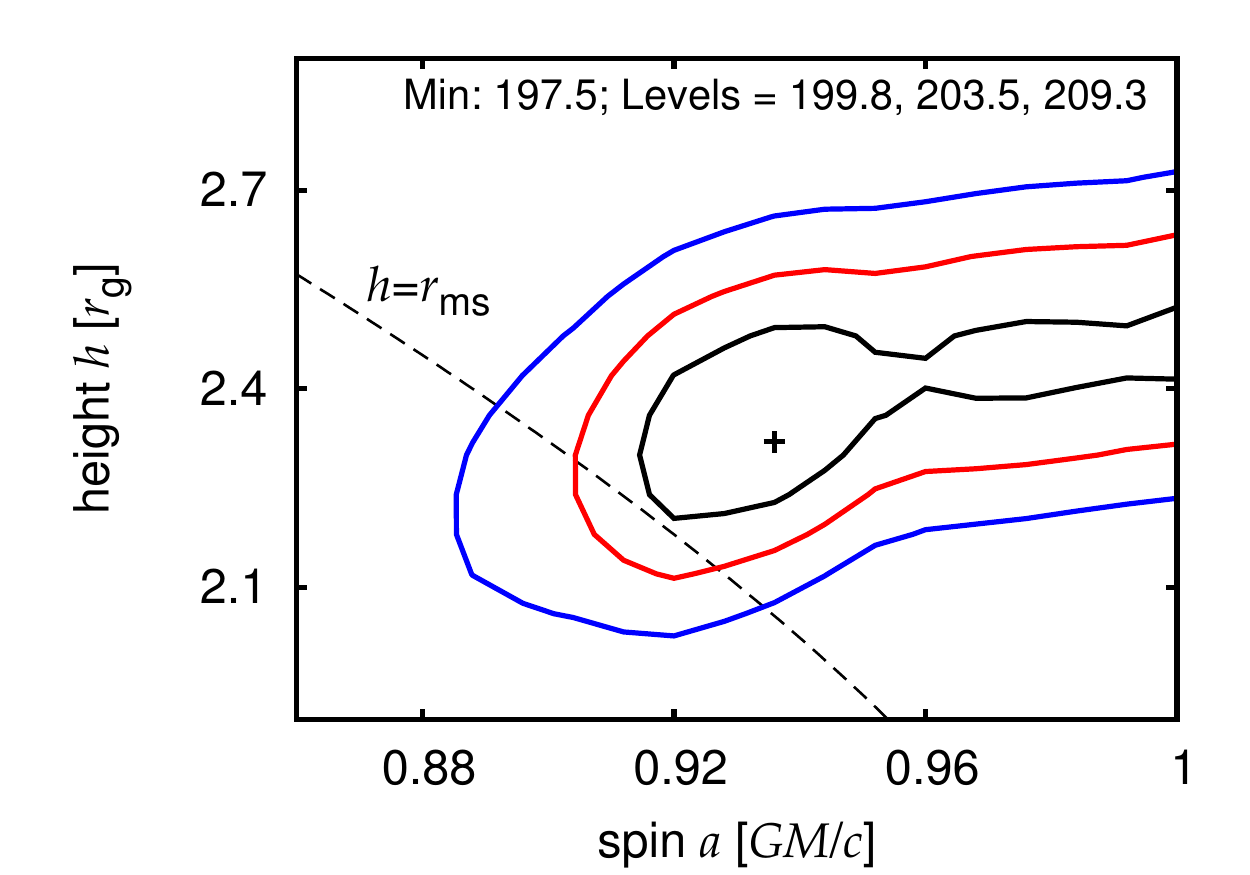}
}
\parbox[t]{0.4\textwidth}{\vspace*{-4cm}
\begin{tabular}{l c c}
Parameter        & Unit      & Best fit value\\
\noalign{\smallskip}
\hline
\noalign{\medskip}
\hspace*{5mm}$a$              &  $GM/c$   &   $0.93_{-0.02}^{+0.02}$\\[3mm]
\hspace*{5mm}$\theta_{\rm o}$ &  deg      &   $21.5_{-0.1}^{+0.6}$\\[3mm]
\hspace*{5mm}$E_{\rm rest}$   &  keV      &   $6.81_{-0.02}^{+0.02}$\\[3mm]
\hspace*{5mm}$h$         &  $GM/c^2$ &   $2.3_{-0.1}^{+0.2}$
\end{tabular}
}
\caption{{\em Left:} The $\chi^2$ contour graphs for the height, $h$, versus 
spin, $a$. Other parameters were kept frozen. {\em Right:} The best fit values 
and their errors for the parameters of the model.}
\label{fig:contours-fit}
\end{figure}

Using the computations from previous sections we have prepared a new XSPEC model 
for the relativistically broadened Iron line in the lamp-post geometry, see the 
Appendix \ref{app:ky} for more details. To find out what value of the height of 
the primary source one can expect in real observations where large spin have 
been observed in the past, we applied our new lamp-post model to the XMM-Newton 
spectrum of a nearby Seyfert 1 galaxy MCG-6-30-15. Very broad iron line was 
reported in this source by several authors 
\citep[e.g.][]{Fabian2002, Ballantyne2003, Vaughan2004, Brenneman2006a}. 
We followed the analysis 
presented in \citet{Svoboda2009} and we employed the same model for the 
underlying X-ray continuum. However, we have used the new KYNRLPLI model instead 
of the KYRLINE \citep{Dovciak2004c} so that we replaced the broken power-law 
radial emissivity by the one
that corresponds to the lamp-post geometry. In XSPEC syntax the overall model 
reads: {\tt PHABS*(POWERLAW+ZGAUSS+ZGAUSS+KYNRLPLI)}. The best-fit parameter 
values and their errors are shown in the table in the right panel of 
Fig.~\ref{fig:contours-fit}. The parameters not 
shown in the table were frozen to their best-fit values from the previous model 
set-up. The reduced $\chi^2$ value was $1.33$.
The contour plot of the primary source height versus the black hole spin is 
shown in the left panel of Fig.~\ref{fig:contours-fit}. The best-fit value for 
the height, $h=2.3\,GM/c^2$, confirms our findings that 
if the primary source of power-law radiation is static, it has to be located 
very close to the black hole so that it illuminates the inner regions by large 
enough intensity to reveal the imprints of high spin in the observed spectrum.

\section{Conclusions}

In this paper we have compared two types of iron line radial emissivity 
profiles, the one governed by the illumination in the lamp-post geometry and 
the radial broken power-law emissivity. We find that
\begin{itemize}
 \item for the primary source height $h\gtrsim3\,GM/c^2$ the lamp-post geometry
      is very well approximated with the broken power-law emissivity with the
      inner power-law index $q_{\rm in}\lesssim 4$ and the outer index
      $q_{\rm out}=3$,
 \item a very high radial power-law index, $q>5$, may be achieved in the
      lamp-post geometry only for very small heights, $h\lesssim 2\,GM/c^2$,
      and, the difference in the line shape in the lamp-post
      geometry and the broken power-law emissivity becomes large,
 \item very high $q$ values originate very close to the central black hole,
      thus it can occur only in the case of a highly spinning black hole,
 \item high $q$ values are mainly due to the gravitational redshift for the
      primary emission with the spectral index $\Gamma > 1$ and due to
      the gravitational aberration for very small heights;
      the contribution of the light bending, as defined in this paper, is 
      moderate.
\end{itemize}

Further we have investigated how the numerically computed emission 
directionality changes the profile of the iron line approximated by isotropic
emission. We show that
\begin{itemize}
 \item the emission from the disc where the incident and emission angles are 
      large is greatly enhanced (limb brightening effect),
 \item the local emission directionality changes the shape of the broad line 
      significantly, however, only for small heights, $h\lesssim 10\,$GM/c$^2$.
\end{itemize}

To summarise our modelling we conclude that in the lamp-post geometry with a
corona approximated by a static isotropic point source a very broad iron line 
profile arises for highly spinning black holes only for the heights 
$h\lesssim 5\,GM/c^2$, while for the heights $h\gtrsim 10\,GM/c^2$ 
the non-spinning and extremely spinning black holes are indistinguishable.

Similar conclusions were drawn by \cite{Dauser2013} for a moving elongated 
jet-like structure along the axis. Another interesting conclusion in their paper 
is that such a vertically extended region may be very well approximated by a 
point source at some effective intermediate height. On the other hand, 
\cite{Wilkins2012} show that such steep emissivities may still be reached even
if the corona is extended horizontally (as far as $30\,GM/c^2$), provided it is
very low above the disc (as low as $2\,GM/c^2$). For a more detailed discussion
on the prospects of spin determination using X-ray reflection we refer the 
reader to a recent paper by \cite{Fabian2014}.

\ack
The research leading to these results has received funding from the European 
Union Seventh Framework Programme (FP7/2007-2013) under grant agreement 
n$^\circ$312789. RG would like to thank French GdR PCHE and the CNRS-AV exchange 
programme for their support. 
GM acknowledges financial support from Agenzia Spaziale Italiana (ASI).

\bibliography{\jobname}

\clearpage
\appendix
\vspace*{-1.5cm}
\section{The radial illumination profile}
\label{app:N_inc}
\vspace*{-1mm}
\includegraphics[width=\textwidth]{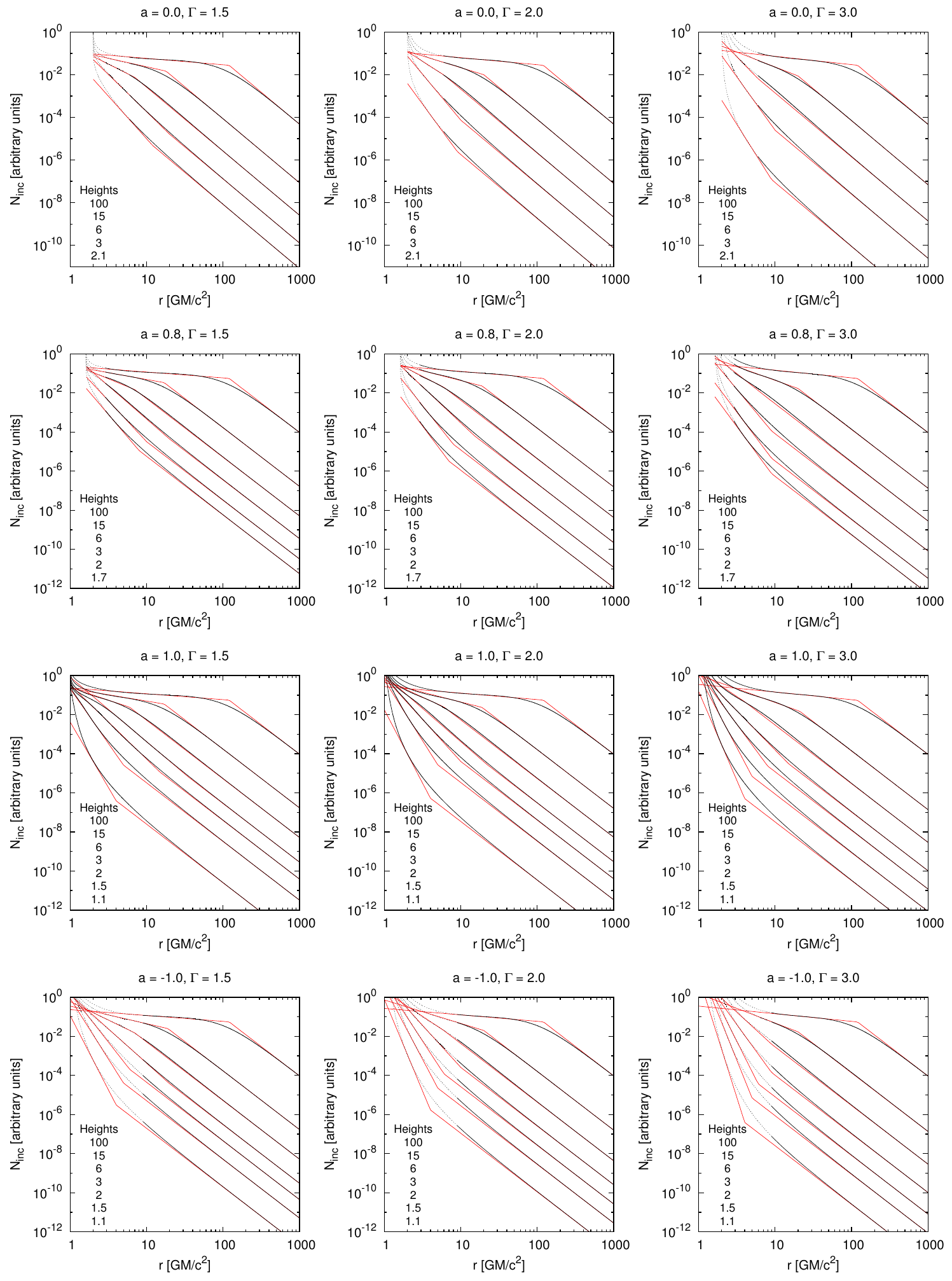}\vspace*{-1cm}
\begin{figure}[hb]
\centering
\caption{
The radial profile of the incident flux, $N_\inc(r)$, defined in 
eq.~(\ref{eq:N_inc1}), for the photon index $\Gamma=1.5,\,2$ and $3$ (left to 
right) and the BH spin $a=0,\,0.8,\,1$ and $-1\,GM/c$ (top to 
bottom). For better clarity, the results shown for different heights, as 
depicted in each panel, are renormalized so as not to cross.
The red lines represent the approximating broken-power-law profiles with 
the outer slope set to $-3$ (see the Tab.~\ref{tab:q1}-\ref{tab:q4} for 
details).}
\vspace*{-1cm}
\label{fig:Ninc}
\end{figure}
\clearpage

\begin{figure}[h]
\centering
\includegraphics[width=\textwidth]{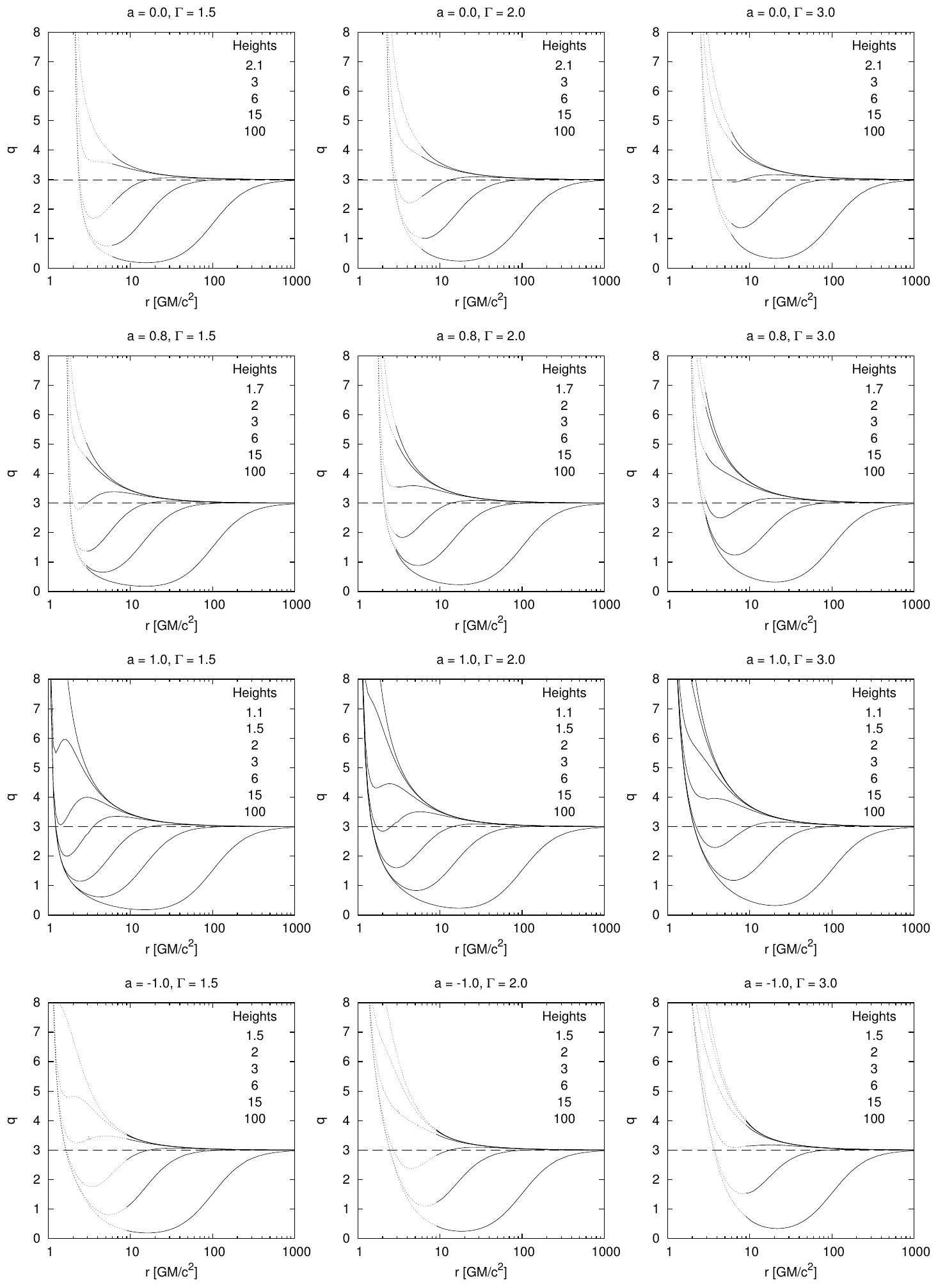}
\caption{The radial dependence of the power-law index $q(r)$, defined in
eq.~(\ref{eq:q}), for the photon index $\Gamma=1.5,\,2$ and $3$ (left to right)
and the BH spin $a=0,\,0.8,\,1$ and $-1\,GM/c$ (top to bottom).
The results for different heights, as depicted on each panel, are shown by solid
lines above and by dotted lines below the marginally stable orbit.}
\label{fig:power}
\end{figure}
\clearpage

\begin{table}[h]
\centering
\begin{tabular}{c@{\hspace*{7mm}}c@{\hspace*{2mm}}c@{\hspace*{7mm}}c
@{\hspace*{2mm}}c@{\hspace*{7mm}}c@{\hspace*{2mm}}c}
\multicolumn{7}{c}{$a=0$}\\[0.5mm]
\hline
\noalign{\smallskip}
$\Gamma$ & \multicolumn{2}{c}{1.5\hspace*{7mm}} & 
\multicolumn{2}{c}{2\hspace*{7mm}} & \multicolumn{2}{c}{3\hspace*{2mm}}\\[0.5mm]
\hline
\noalign{\smallskip}
$h$ & $q_{\rm i}$ & $r_{\rm b}$ & $q_{\rm i}$ & $r_{\rm b}$ & $q_{\rm i}$ & 
$r_{\rm b}$\\
\noalign{\smallskip}
\hline
\noalign{\smallskip}
100 & 0.3 & 120 & 0.33 & 120 & 0.39 & 120 \\[1mm]
15  & 0.8 & 18  & 1.1 & 20   & 1.4 & 20   \\[1mm]
6   & 1.8 & 7   & 2.3 & 6    & 3.7 & 6    \\[1mm]
3   & 3.5 & 20  & 4.3 & 10   & 5.0 & 10   \\[1mm]
2.1 & 4.0 & 12  & 4.9 & 9    & 5.7 & 9
\end{tabular}
\caption{The values of the inner slope for the broken power-law, $q_{\rm i}$, 
and the break radius, $r_{\rm b}$, for different height, $h$, (rows) and photon 
index, $\Gamma$, (columns) in the case of a non-rotating Schwarzschild black hole
(with the spin $a=0\,GM/c$, horizon $r_{\rm h}=2\,GM/c^2$ and marginally
stable orbit $r_{\rm ms}=6\,GM/c^2$). Both the height and the break radius are 
specified in units of $GM/c^2$. These values correspond to the broken
power-law dependences in the top panels in Fig.~\ref{fig:Ninc}.}
\label{tab:q1}
\end{table}

\bigskip
\begin{table}[h]
\centering
\begin{tabular}{c@{\hspace*{7mm}}c@{\hspace*{2mm}}c@{\hspace*{7mm}}
c@{\hspace*{2mm}}c@{\hspace*{7mm}}c@{\hspace*{2mm}}c}
\multicolumn{7}{c}{$a=0.8$}\\[0.5mm]
\hline
\noalign{\smallskip}
$\Gamma$ & \multicolumn{2}{c}{1.5\hspace*{7mm}} & 
\multicolumn{2}{c}{2\hspace*{7mm}} & \multicolumn{2}{c}{3\hspace*{2mm}}\\[0.5mm]
\hline
\noalign{\smallskip}
$h$ & $q_{\rm i}$ & $r_{\rm b}$ & $q_{\rm i}$ & $r_{\rm b}$ & $q_{\rm i}$ & 
$r_{\rm b}$\\
\noalign{\smallskip}
\hline
\noalign{\smallskip}
100 & 0.3 & 120 & 0.33 & 120  & 0.39 & 120 \\[1mm]
15  & 0.7 & 17  & 1.0  & 19   & 1.4  & 20  \\[1mm]
6   & 1.6 & 7   & 2.0  & 6    & 3.1  & 8   \\[1mm]
3   & 3.4 & 20  & 3.7  & 15   & 4.7  & 9   \\[1mm]
2   & 4.2 & 10  & 4.9  & 8    & 5.2  & 9   \\[1mm]
1.7 & 4.6 & 8   & 5.2  & 7    & 5.3  & 9
\end{tabular}
\caption{The same as in Table~\ref{tab:q1} but for the co-rotating Kerr
black hole with the spin $a=0.8\,GM/c$ (horizon $r_{\rm h}=1.6\,GM/c^2$ and
marginally stable orbit $r_{\rm ms}=2.9\,GM/c^2$). These values correspond to
the broken power-law dependences in the second row panels in
Fig.~\ref{fig:Ninc}.}
\label{tab:q2}
\end{table}

\clearpage
\begin{table}[h]
\centering
\begin{tabular}{c@{\hspace*{7mm}}c@{\hspace*{2mm}}c@{\hspace*{7mm}}
c@{\hspace*{2mm}}c@{\hspace*{7mm}}c@{\hspace*{2mm}}c}
\multicolumn{7}{c}{$a=1$}\\[0.5mm]
\hline
\noalign{\smallskip}
$\Gamma$ & \multicolumn{2}{c}{1.5\hspace*{7mm}} & \multicolumn{2}{c}{2\hspace*{7mm}} & 
\multicolumn{2}{c}{3\hspace*{2mm}}\\[0.5mm]
\hline
\noalign{\smallskip}
$h$ & $q_{\rm i}$ & $r_{\rm b}$ & $q_{\rm i}$ & $r_{\rm b}$ & $q_{\rm i}$ & $r_{\rm b}$\\
\noalign{\smallskip}
\hline
\noalign{\smallskip}
100 & 0.3  & 120 & 0.33 & 120 & 0.39 & 120  \\[1mm]
15  & 0.65 & 17  &  1.0 & 19  & 1.4  & 21   \\[1mm]
6   & 1.3  & 6   &  1.8 & 6   & 2.6  & 4    \\[1mm]
3   & 3.3  & 35  &  3.4 & 22  & 4.0  & 15   \\[1mm]
2   & 3.8  & 15  &  4.3 & 10  & 5.3  & 8    \\[1mm]
1.5 & 5.6  & 5   &  6.2 & 5   & 7.4  & 5    \\[1mm]
1.1 & 6.7  & 4   &  7.5 & 4   & 9.1  & 4
\end{tabular}
\caption{The same as in Table~\ref{tab:q1} but for the extreme co-rotating Kerr
black hole (with the spin $a=1\,GM/c$, horizon $r_{\rm h}=1\,GM/c^2$ and
marginally stable orbit $r_{\rm ms}=1\,GM/c^2$). These values correspond to
the broken power-law dependences in the third row panels in
Fig.~\ref{fig:Ninc}.}
\label{tab:q3}
\end{table}

\begin{table}[h]
\centering
\begin{tabular}{c@{\hspace*{7mm}}c@{\hspace*{2mm}}c@{\hspace*{7mm}}
c@{\hspace*{2mm}}c@{\hspace*{7mm}}c@{\hspace*{2mm}}c}
\multicolumn{7}{c}{$a=-1$}\\[0.5mm]
\hline
\noalign{\smallskip}
$\Gamma$ & \multicolumn{2}{c}{1.5\hspace*{7mm}} & \multicolumn{2}{c}{2\hspace*{7mm}} & 
\multicolumn{2}{c}{3\hspace*{2mm}}\\[0.5mm]
\hline
\noalign{\smallskip}
$h$ & $q_{\rm i}$ & $r_{\rm b}$ & $q_{\rm i}$ & $r_{\rm b}$ & $q_{\rm i}$ & $r_{\rm b}$\\
\noalign{\smallskip}
\hline
\noalign{\smallskip}
100 & 0.3 & 120 & 0.33 & 120 &  0.39 & 120 \\[1mm]
15  & 0.9 & 19  & 1.2  & 20  &  1.5  & 20  \\[1mm]
6   & 1.9 & 7   & 2.6  & 6   &  3.6  & 9   \\[1mm]
3   & 3.5 & 20  & 4.3  & 10  &  5.4  & 9   \\[1mm]
2   & 5.0 & 6   & 6.0  & 6   &  7.8  & 6   \\[1mm]
1.5 & 6.2 & 5   & 7.4  & 5   &  9.7  & 5   \\[1mm]
1.1 & 7.6 & 4   & 9.4  & 4   & 12.4  & 4
\end{tabular}
\caption{The same as in Table~\ref{tab:q1} but for the extreme counter-rotating
Kerr black hole (with the spin $a=-1\,GM/c$, horizon $r_{\rm h}=1\,GM/c^2$ and
marginally stable orbit $r_{\rm ms}=9\,GM/c^2$). These values correspond to the
broken power-law dependences in the bottom panels in Fig.~\ref{fig:Ninc}.}
\label{tab:q4}
\end{table}

\clearpage
\section{Maps of the transfer function, emission directionality and observed 
flux}
\label{app:maps}

\begin{figure}[h]
\centering
\includegraphics[width=\textwidth]{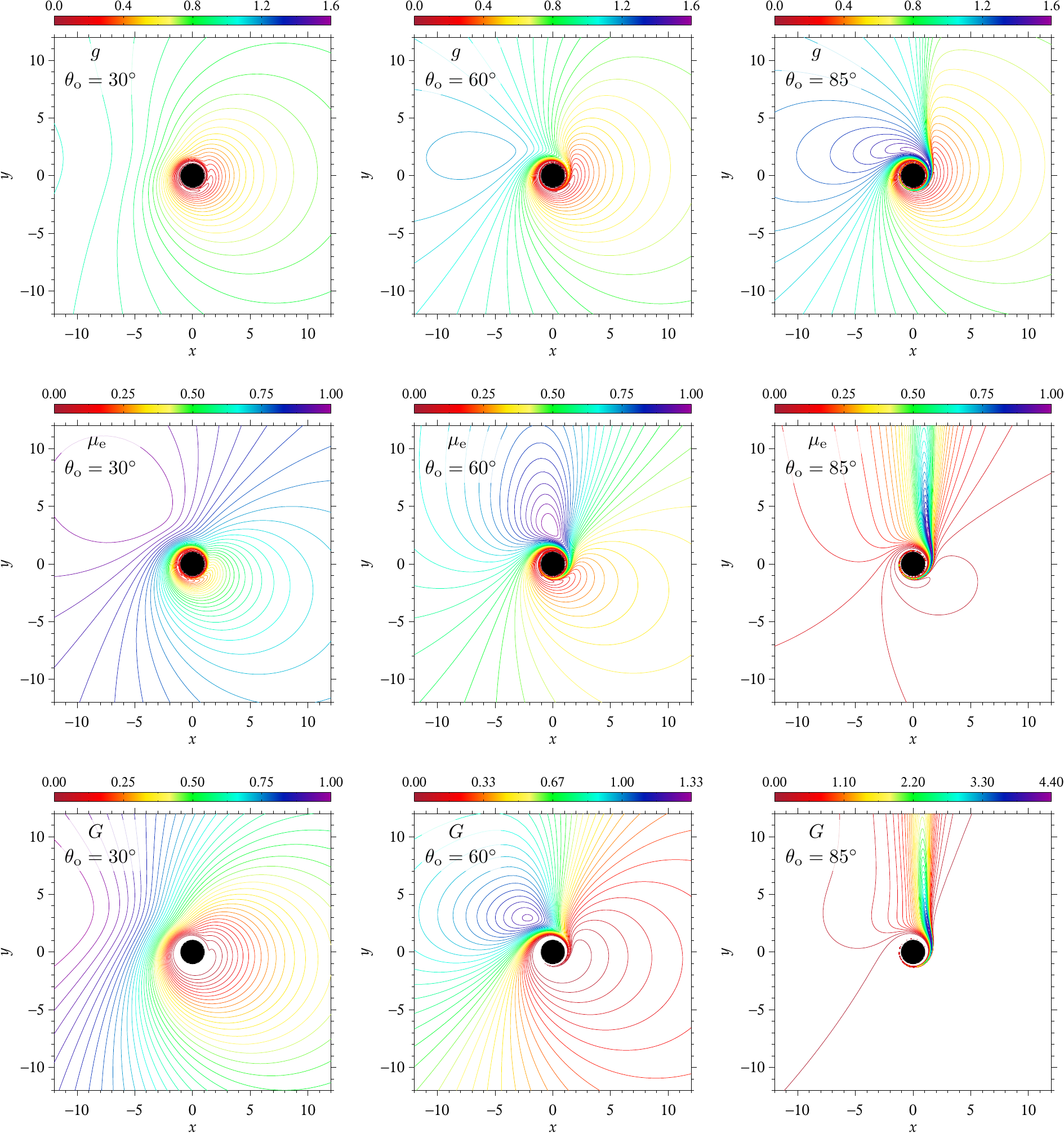}
\caption{\label{fig:gcosG_100} 
The equatorial plane map of the energy shift, $g$, cosine of emission angle, 
$\mu_{\rm e}$, and transfer function, $G$, (top to bottom) 
for the co-rotating Kerr black hole ($a=1\,GM/c$) and three inclination angles, 
$\theta_{\rm o}=30^\circ,\,60^\circ$ and $85^\circ$ (left to right).}
\end{figure}

\begin{figure}
\centering
\includegraphics[width=\textwidth]{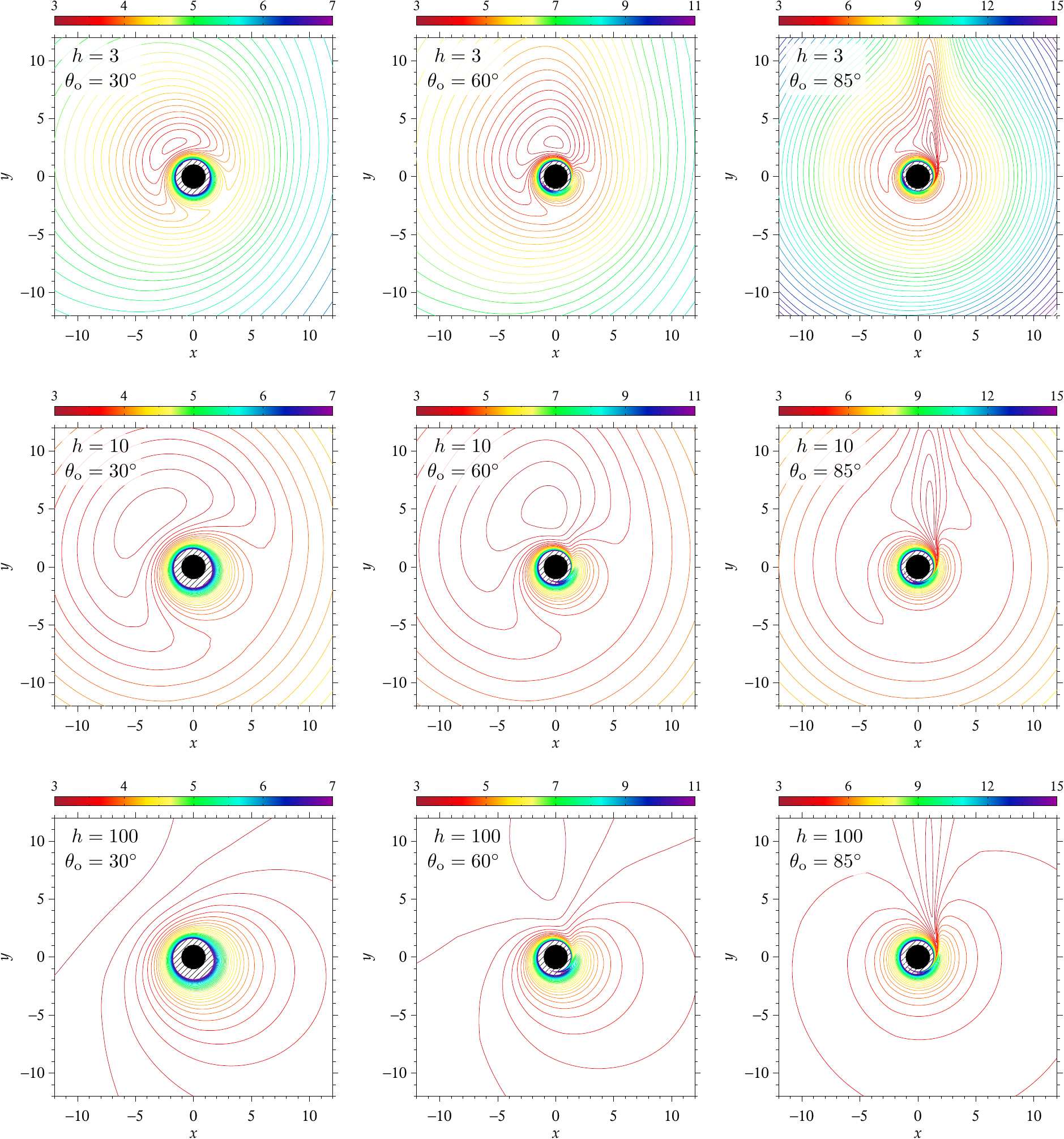}
\caption{\label{fig:M_100}
The equatorial plane map of the local flux emission directionality, 
$\cal{M}(\mu_{\rm i}, \mu_{\rm e})$, 
for the co-rotating Kerr black hole ($a=1\,GM/c$) and three inclination angles, 
$\theta_{\rm o}=30^\circ,\,60^\circ$ and $85^\circ$ (left to right), and three
heights of the primary source, $h=3,\,10$ and $100\,GM/c^2$ (top to bottom). }
\end{figure}

\begin{figure}
\centering
\includegraphics[width=\textwidth]{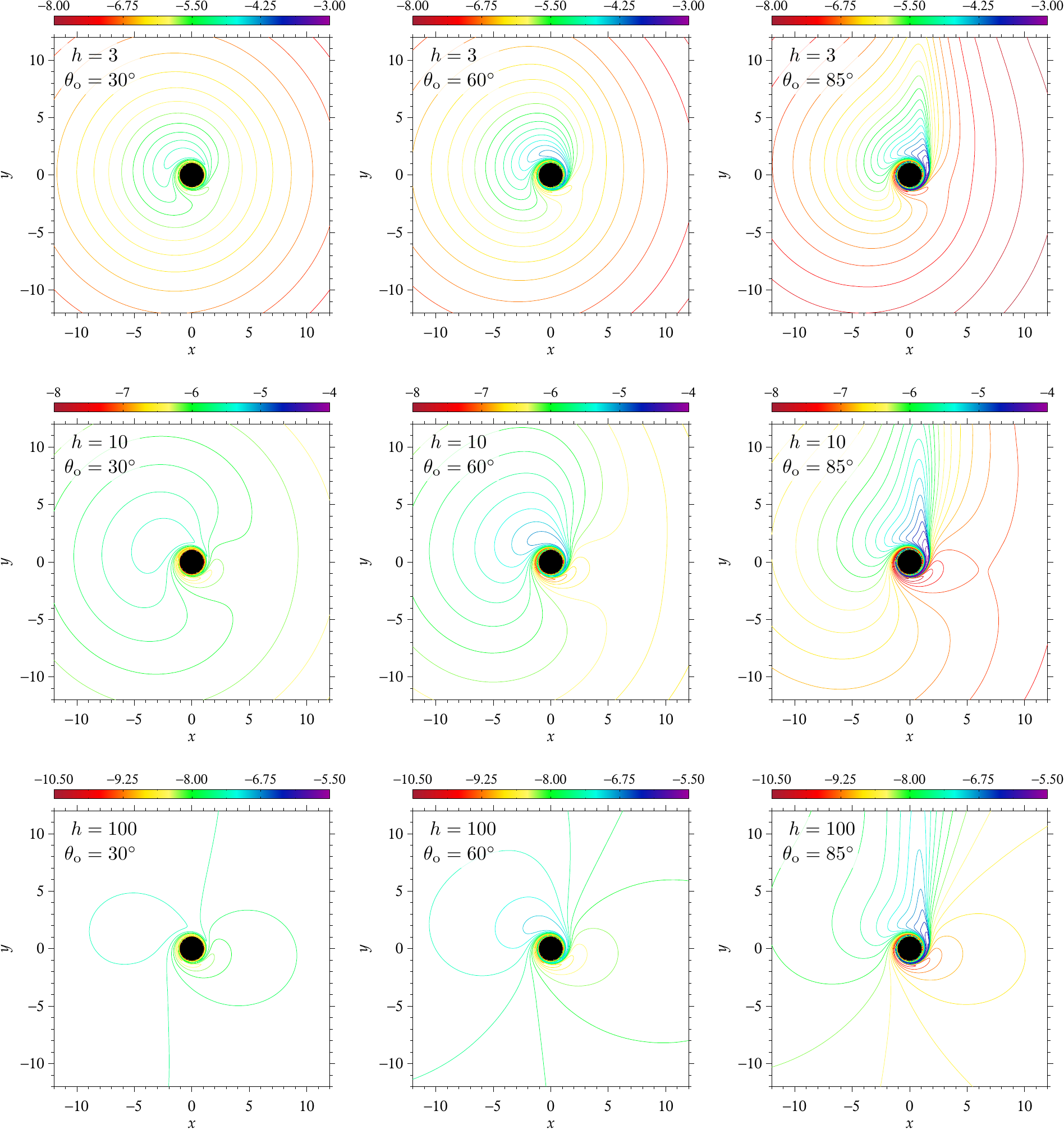}
\caption{\label{fig:F_100}
The equatorial plane map of the observed line flux, $F_{\rm obs}(r, \varphi)$, 
for the co-rotating Kerr black hole ($a=1\,GM/c$) and three inclination angles, 
$\theta_{\rm o}=30^\circ,\,60^\circ$ and $85^\circ$ (left to right), and three
heights of the primary source, $h=3,\,10$ and $100\,GM/c^2$ (top to bottom). }
\end{figure}

\begin{figure}
\centering
\includegraphics[width=\textwidth]{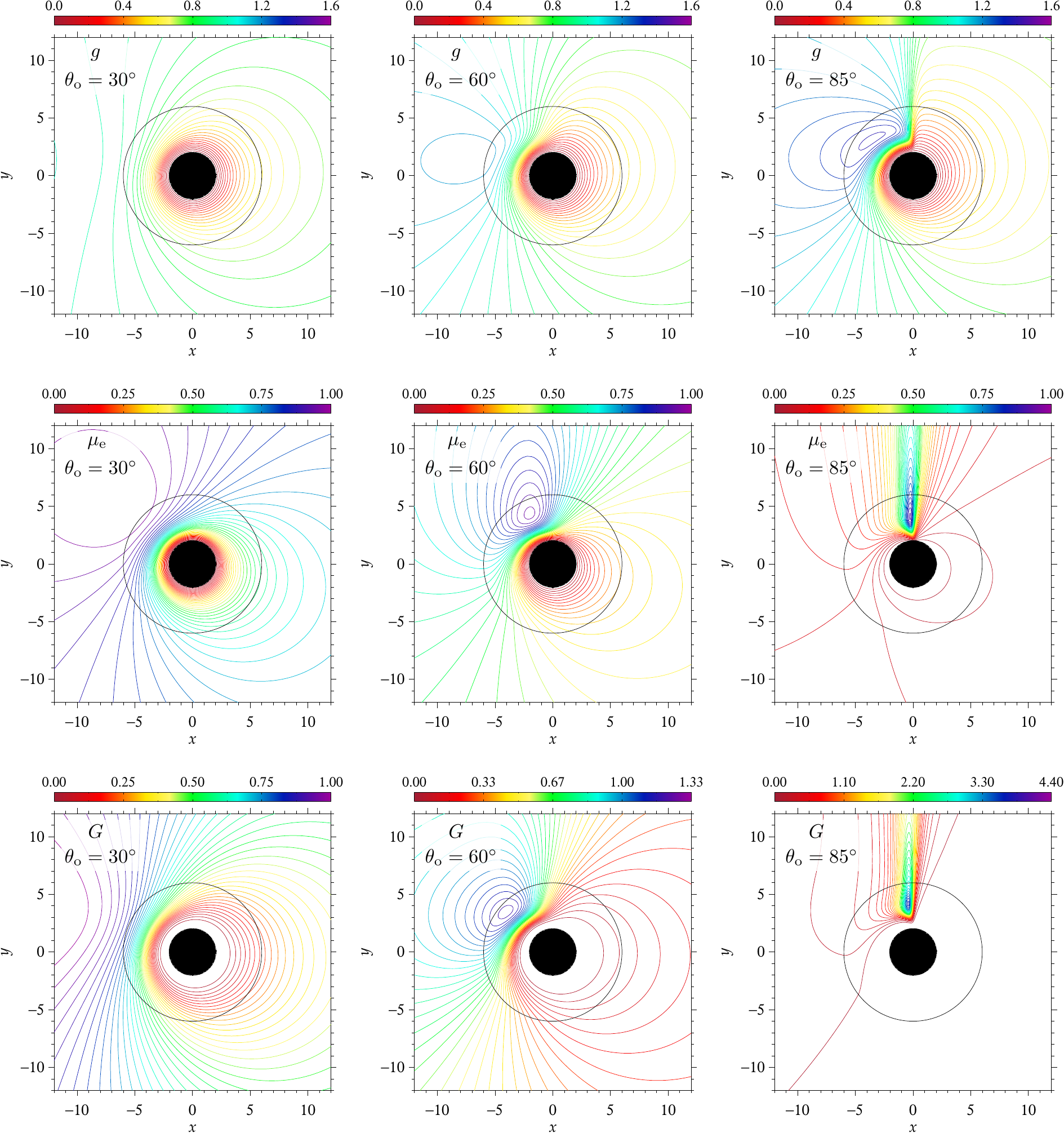}
\caption{\label{fig:gcosG_200} 
The equatorial plane map of the energy shift, $g$, cosine of emission angle, 
$\mu_{\rm e}$, and transfer function, $G$, (top to bottom) 
for the Schwarzschild black hole ($a=0\,GM/c$) and three inclination angles, 
$\theta_{\rm o}=30^\circ,\,60^\circ$ and $85^\circ$ (left to right).
The marginally stable orbit at $r_{\rm ms}=6\,$GM/c$^2$ is denoted by a black 
circle.}
\end{figure}

\begin{figure}
\centering
\includegraphics[width=\textwidth]{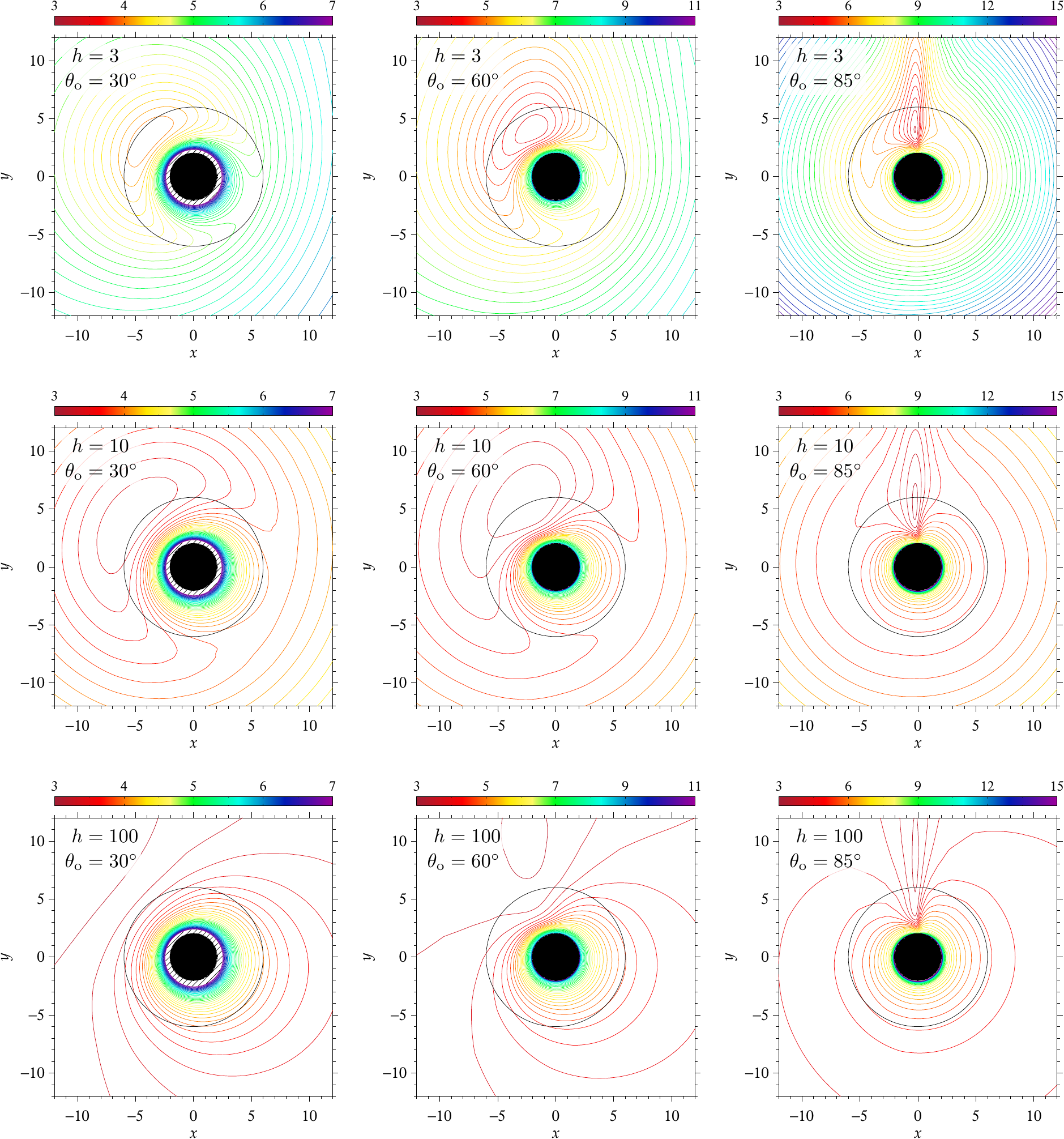}
\caption{\label{fig:M_200}
The equatorial plane map of the local flux emission directionality, 
$\cal{M}(\mu_{\rm i}, \mu_{\rm e})$, 
for the Schwarzschild black hole ($a=0\,GM/c$) and three inclination angles, 
$\theta_{\rm o}=30^\circ,\,60^\circ$ and $85^\circ$ (left to right), and three
heights of the primary source, $h=3,\,10$ and $100\,GM/c^2$ (top to bottom). 
The marginally stable orbit at $r_{\rm ms}=6\,$GM/c$^2$ is denoted by a black 
circle.}
\end{figure}

\begin{figure}
\centering
\includegraphics[width=\textwidth]{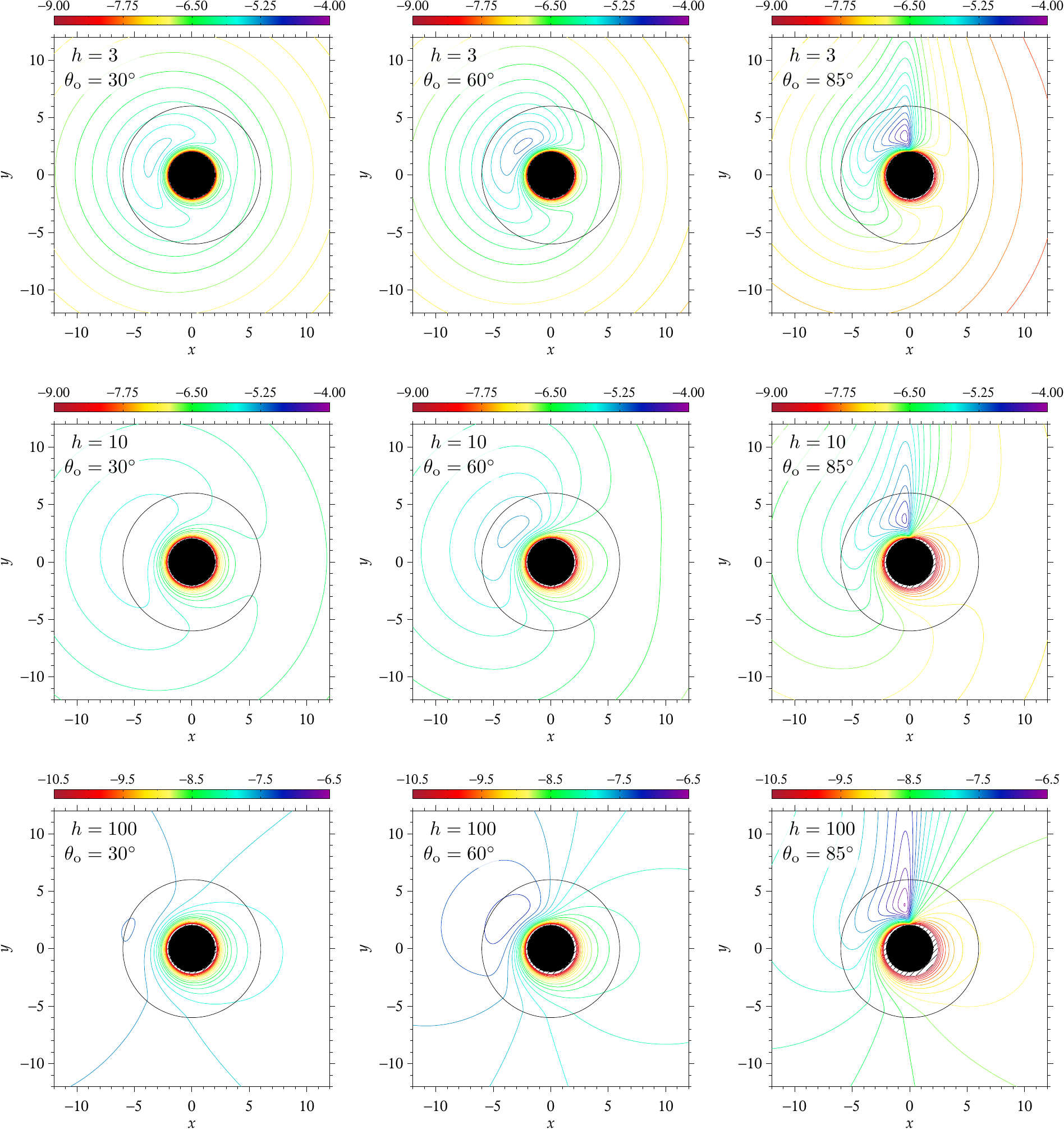}
\caption{\label{fig:F_200} 
The equatorial plane map of the observed line flux, $F_{\rm obs}(r, \varphi)$, 
for the Schwarzschild black hole ($a=0\,GM/c$) and three inclination angles, 
$\theta_{\rm o}=30^\circ,\,60^\circ$ and $85^\circ$ (left to right), and three
heights of the primary source, $h=3,\,10$ and $100\,GM/c^2$ (top to bottom).
The marginally stable orbit at $r_{\rm ms}=6\,$GM/c$^2$ is denoted by a black 
circle.}
\end{figure}

\begin{figure}
\centering
\includegraphics[width=\textwidth]{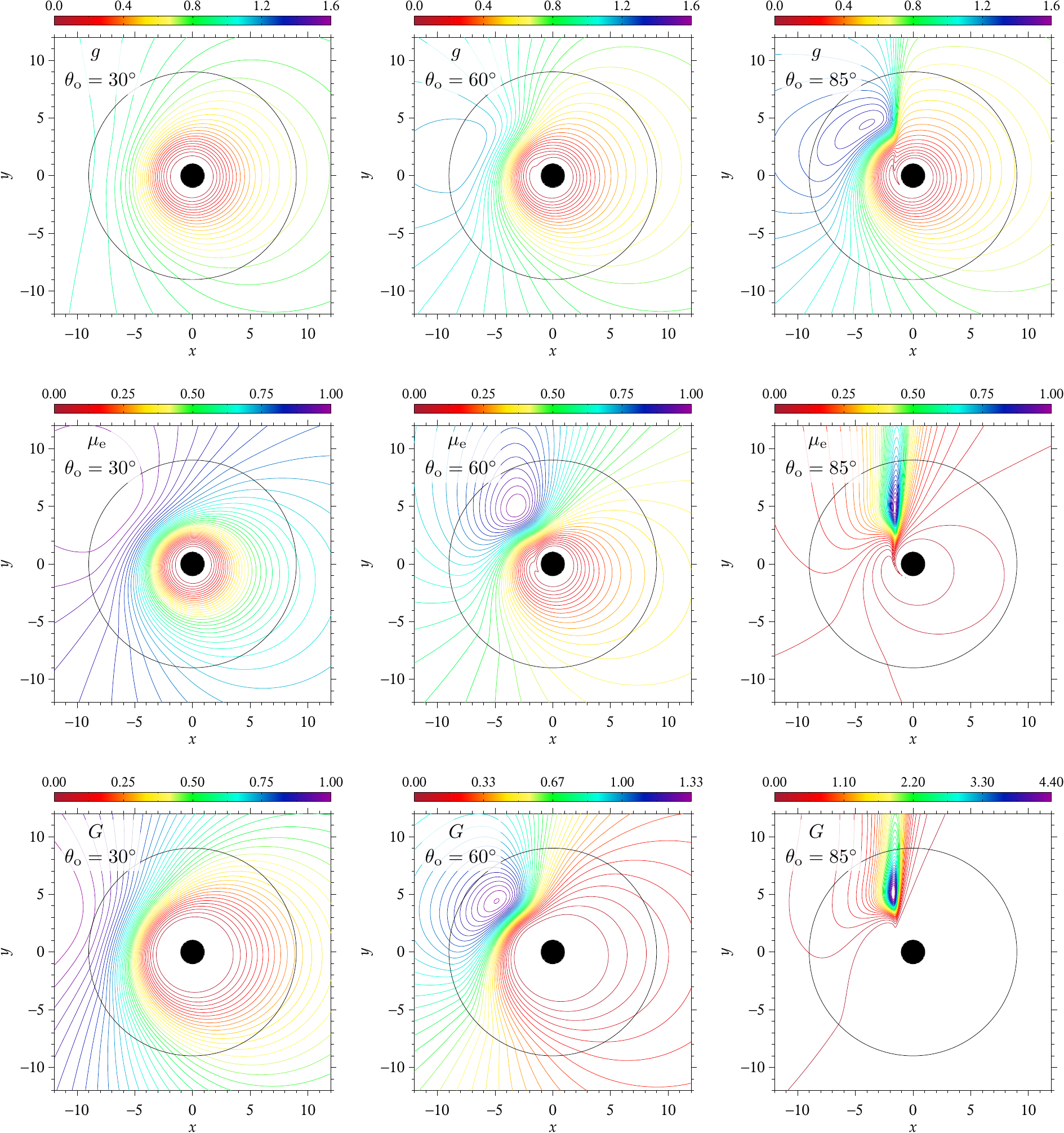}
\caption{\label{fig:gcosG_300}
The equatorial plane map of the energy shift, $g$, cosine of emission angle, 
$\mu_{\rm e}$, and transfer function, $G$, (top to bottom) 
for the counter-rotating Kerr black hole ($a=-1\,GM/c$) and three inclination
angles, $\theta_{\rm o}=30^\circ,\,60^\circ$ and $85^\circ$ (left to right).
The marginally stable orbit at $r_{\rm ms}=9\,$GM/c$^2$ is denoted by a black 
circle.}
\end{figure}

\begin{figure}
\centering
\includegraphics[width=\textwidth]{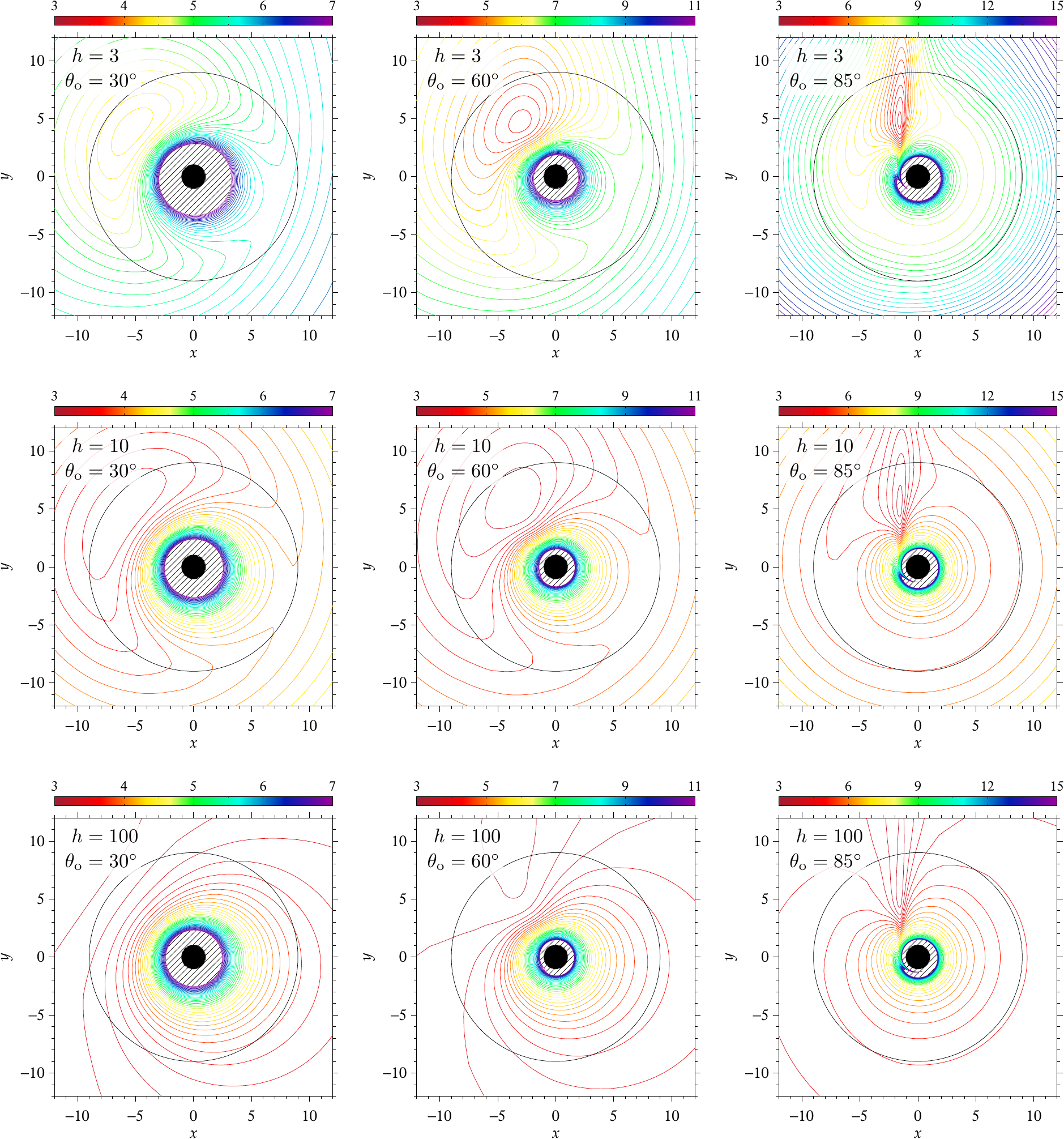}
\caption{\label{fig:M_300}
The equatorial plane map of the local flux emission directionality, 
$\cal{M}(\mu_{\rm i}, \mu_{\rm e})$, 
for the counter-rotating Kerr black hole ($a=-1\,GM/c$) and three inclination 
angles, $\theta_{\rm o}=30^\circ,\,60^\circ$ and $85^\circ$ (left to right), and 
three heights of the primary source, $h=3,\,10$ and $100\,GM/c^2$ (top to 
bottom). The marginally stable orbit at $r_{\rm ms}=9\,$GM/c$^2$ is denoted by 
a black circle.}
\end{figure}

\begin{figure}
\centering
\includegraphics[width=\textwidth]{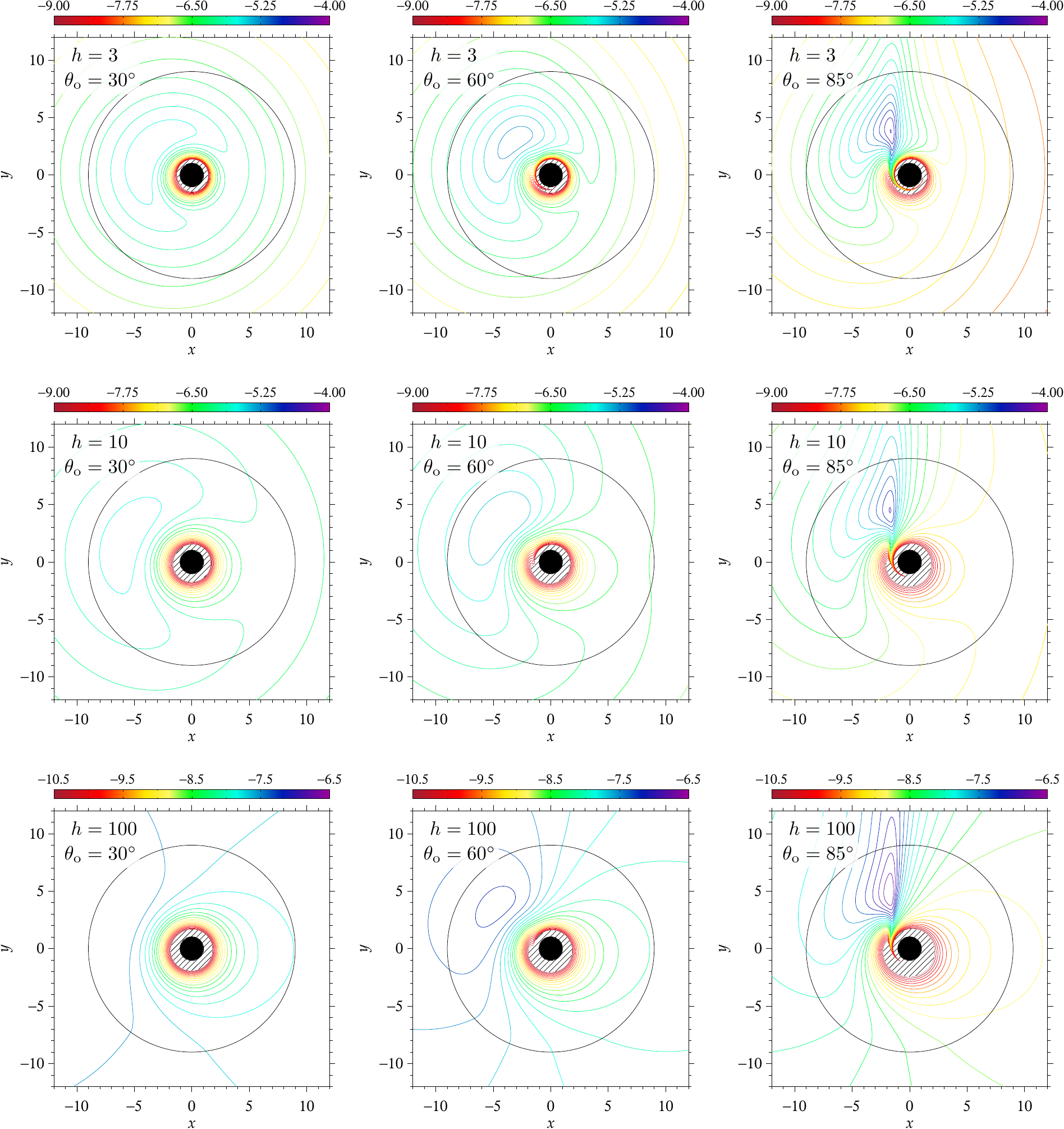}
\caption{\label{fig:F_300} 
The equatorial plane map of the observed line flux, $F_{\rm obs}(r, \varphi)$, 
for the counter-rotating Kerr black hole ($a=-1\,GM/c$) and three inclination 
angles, $\theta_{\rm o}=30^\circ,\,60^\circ$ and $85^\circ$ (left to right), and 
three heights of the primary source, $h=3,\,10$ and $100\,GM/c^2$ (top to 
bottom). The marginally stable orbit at $r_{\rm ms}=9\,$GM/c$^2$ is denoted by 
a black circle.}
\end{figure}

\clearpage
\section{The relativistic iron line model for the lamp-post geometry}
\label{app:ky}

To be able to use the lamp-post scheme with the data we have developed the model
for XSPEC \citep{Arnaud1996} -- KYNRLPLI (KY Non-axisymmetric Relativistic 
Lamp-Post LIne). This model is based on the non-axisymmetric version of the KY 
package of models \citep{Dovciak2004,Dovciak2004c,Dovciak2004d}.

The model approximates the corona above the disc by a static isotropic point 
source located on the rotational axis at height, $h$, above the disc (measured
from the centre of the black hole). Thus the radial emissivity profile is given 
by the illumination from such corona. All relativistic effects are taken into 
account all the way from the primary source to the disc and from the disc to the 
observer.

The local flux angular dependence, $\cal{M}(\mu_{\rm i}, \mu_{\rm e})$, is 
computed with the Monte Carlo code NOAR \citep{Dumont2000}, see also 
Fig.~\ref{fig:local_flux} and Appendix~\ref{app:maps}.

As is usual in non-axisymmetric KY models, it is possible to choose that the 
radiation comes only from a segment of the disc to simulate an emission from a 
spot. The inner and outer radius might be set either in physical units of 
$GM/c^2$ or as a multiple of the marginally stable orbit, $r_{\rm ms}$.

On the other hand we have added a possibility to obscure part of the disc by 
a circular cloud in the observer's sky (i.e. farther away from the centre). 
The centre of the cloud is set in impact parameters, $\alpha$ and $\beta$, where
$\alpha$ is positive for approaching side of the disc and $\beta$ is positive
above the black hole and negative below it (in the observer's sky).

\begin{table}[bh]
\centering
\begin{tabular}{lcccl}
param. & param. & unit   & possible & description \\
number &        &        & values   &             \\
\noalign{\smallskip}
\hline\hline
\noalign{\smallskip}
par1   & $a$    & $GM/c$ & -1 -- 1  & black hole angular momentum\\[1mm]
par2   & $\theta_{\rm o}$ & deg & 0 -- 89 & observer inclination \\
       &        & &        & ($0^\circ$ -- pole, $90^\circ$ -- disc)\\[1mm]
par3   & $r_{\rm in}$ & $GM/c^2$ & 1 -- 1000 & inner disc edge\\[1mm]
par4   & ms       & & 0, 1, 2 & changes definition of inner edge\\
       &          & &         & 0:  $r_{\rm in}$ = par3 \\
       &          & &         & 1:  $r_{\rm in}$ = par3 but if 
                                      par3$\,<\,r_{\rm ms}$\\
       &          & &         & \phantom{1:}  then $r_{\rm in} = r_{\rm ms}$\\
       &          & &         & 2: $r_{\rm in}$ = par3$\times r_{\rm ms}$,
                                  $r_{\rm out}$ = par5$\times r_{\rm ms}$\\[1mm]
par5   & $r_{\rm out}$ & $GM/c^2$ & 1 -- 1000 & outer disc edge\\[1mm]
par6   & $\varphi_{\rm o}$     & deg & -180 -- 180 & lower azimuth of the disc segment\\[1mm]
par7   & $\Delta\varphi$ & deg & 0 -- 360    & width of the disc segment\\[1mm]
par8   & $h$ & $GM/c^2$ & 1 -- 100 & height (location) of the primary\\[1mm]
par9   & $\Gamma$ &  & 1.1 -- 3 & primary energy power-law index\\[1mm]
\end{tabular}
\caption{Description of the KYNRLPLI parameters par1 -- par9.}
\label{tab:kynrlpli1}
\end{table}

\begin{table}[th]
\begin{tabular}{lcccl}
param. & param. & unit   & possible & description \\
number &        &        & values   &             \\
\noalign{\smallskip}
\hline\hline
\noalign{\smallskip}
par10  & $\alpha_{\rm c}$ & $GM/c^2$ &  & $\alpha$-position of the obscuring 
                                          cloud\\[1mm]
par11  & $\beta_{\rm c}$  & $GM/c^2$ &  & $\beta$-position of the obscuring 
                                          cloud\\[1mm]
par12  & $r_{\rm c}$ & $GM/c^2$ &  & radius of the obscuring cloud\\[1mm]
par13  & zshift & & & overall Doppler shift\\[1mm]
par14  & ntable & & 80 & defines fits file with tables\\[1mm]
par15  & $n_r$ & & 1 -- $10^4$& number of radial grid points\\[1mm]
par16  & division & & 0, 1 & type of step in radial integration\\
       &          & &      & (0 -- equidistant, 1 -- exponential)\\[1mm]
par17  & $n_{\varphi}$ & & 1 -- $2\times10^4$ & number of azimuthal grid 
                                                points\\[1mm]
par18  & smooth & & 0, 1   & smooth the resulting spectrum\\
       &        & &        & (0 -- no, 1 -- yes)\\[1mm]
par19  & Stokes & & 0 -- 6 & output of the computation:\\
       &        & &        & 0: photon number density flux\\
       &        & &        & \phantom{0: }(Stokes parameter I/E)\\
       &        & &        & 1: Stokes parameter Q/E\\
       &        & &        & 2: Stokes parameter U/E\\
       &        & &        & 3: Stokes parameter V/E\\
       &        & &        & 4: degree of polarization\\
       &        & &        & 5: linear polarization angle,\\
       &        & &        & \phantom{5:} $\chi=\frac{1}{2}\,{\rm atan}\,
                                            \frac{U}{Q}$\\
       &        & &        & 6: circular polarization angle,\\
       &        & &        & \phantom{6:} $\psi=\frac{1}{2}\,{\rm asin}\,
                                            \frac{V}{\sqrt{Q^2+U^2+V^2}}$\\[1mm]
par20  & $n_{\rm threads}$ & & 1 -- 100 & number of computation threads\\
\end{tabular}
\caption{Description of the KYNRLPLI parameters par10 -- par20.}
\label{tab:kynrlpli2}
\end{table}

The model can be used also for computing polarisation in a very simple toy model
where all the local line polarisation in the disc is fully polarised 
perpendicularly to the disc.

Due to the fact that the non-axisymmetric models integrate the emission over the
disc and thus are slower, the model may be run in multiple threads to use all
CPU cores available for computing. In this case the XSPEC may need to be run 
with a preloaded thread library 
(e.g. {\tt LD\_PRELOAD=libpthread.so.0 \$HEADAS/bin/xspec})

As usual for spectral line models inside XSPEC, also the KYNRLPLI model is 
normalised to the unit total photon flux.

The model parameters, their definitions and possible values are summarised in 
Tab.~\ref{tab:kynrlpli1} and \ref{tab:kynrlpli2}.


\end{document}